\newcommand\ignore[1]{}
\documentclass[10pt]{article}
\usepackage{amsmath,amssymb,graphicx,url,hyperref}
\usepackage{color}
\usepackage{epstopdf}

\newcommand\be{\begin{equation}}
\newcommand\ee{\end{equation}}
\newcommand\bea{\begin{eqnarray}}
\newcommand\eea{\end{eqnarray}}
\newcommand{\bdm}{\begin{displaymath}}
\newcommand{\edm}{\end{displaymath}}
\newcommand\nn{ \nonumber\\}

\makeatletter
\@addtoreset{equation}{section}
\makeatother

\title{Strong Coupling Expansion for the Conformal  Pomeron/Odderon  Trajectories}

 \author{Richard  C. Brower\footnote{Physics Department,
Boston University, Boston MA 02215},
Miguel S. Costa\footnote{Centro de F\'\i sica do Porto, Departamento de F\'\i sica e Astronomia, Faculdade Ci\^encias da Universidade do Porto}, Marko Djuri\'c\footnotemark[\value{footnote}], \\ 
Timothy Raben\footnote{Physics Department, Brown University,
Providence, RI 02912}, and  Chung-I Tan\footnotemark[\value{footnote}]
}

\begin{document}

\maketitle

\begin{abstract}
From the perspective of AdS/CFT  the Pomeron is identified with a
Reggeized Graviton,   while the Odderons  correspond to  Reggeized anti-symmetric $AdS_5$ Kalb-Ramond tensor-fields.    
In this paper, we consider the strong coupling expansion of the dimension of the leading twist operators dual to these Regge trajectories,  
$\Delta(j)$, to determine its analytic continuation in $j$ beyond the diffusion limit. In particular, we compute the strong coupling expansion of the intercept to order 
$\lambda^{-3}$, where $\lambda$ is the t'Hooft coupling,  for both the  Pomeron, which is   $C=+1$ crossing-even, and the ``Odderons", which are  the leading  $C=-1$ crossing-odd Regge singularities. 
We discuss the spectral curves of the class of single-trace operators to which these string modes couple. 
\end{abstract}

\newpage
\section{Introduction}
\label{sec:intro}

{\it AdS/CFT correspondence}~\cite{Maldacena:1997re,Gubser:1998bc,Witten:1998qj,Witten:1998zw},  has
provided a useful perspective on several  domains of  non-perturbative  QCD, such as heavy ion collisions, low-energy meson dynamics,
and high energy scattering.
In particular,  at strong coupling the leading Pomeron exchange has been identified as a {\it Reggeized $AdS$ Graviton},  in the  planar approximation to 
  ${\cal N}=4$ Super Yang Mills (SYM)
Theory~\cite{Brower:2006ea}.

\begin{figure}[bh]
\begin{center}
\includegraphics[width = 3.0in]{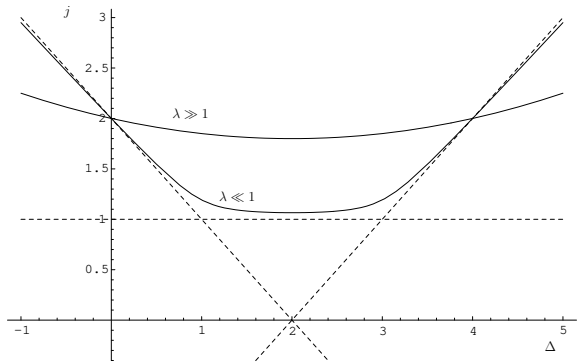}
\caption{The conformal Regge trajectory reproduced from Ref.~\cite{Brower:2006ea} :  Schematic form of the $\Delta-j$ relation for twist-2 spectral curve at  weak ($\lambda\ll 1$) and strong coupling
($\lambda\gg 1$).  
}
\label{fig:BFKLDGLAP}
\end{center}
\end{figure}

A crucial observation made in~\cite{Brower:2006ea} is the role
played by the  analytic continuation in the $\Delta-j$ plane for   anomalous dimensions,
$\gamma(j)= \Delta(j) -j -\tau$, for the leading twist operators as a function of $j$ and the 't Hooft coupling
$\lambda = g^2 N_c$. In a conformal field theory, the inverse curve in the  ``Dimension-$j$" plane, $j(\Delta)$,
  plays a central
role analogous to the traditional Regge pole trajectory $\alpha(t)$  in ``Energy-$j$'' plane.  As a spectral curve,  $\Delta(j)$ has a
remarkable symmetry due to conformal invariance:
the inverse curve, $j(\Delta)$, is symmetric under 
$
\Delta\leftrightarrow 4 - \Delta
$,~\footnote{More
  succinctly stated, conformal symmetry implies a $j$-plane trajectory as a function of 
  $ \Delta (\Delta-4)=M^2_{ads} R^2_{ads}  $ just as Lorentz invariance
  implies Regge  j-plane trajectories as a function of $\alpha' t $. We will occasionally refer to $j(\Delta)$  as the ``conformal Reggeon spin", or simply ``Reggeon spin".}
with a minimum at $\Delta=2$, 
 as shown in
Fig. \ref{fig:BFKLDGLAP} for the leading twist-2 spectral curve. 
 At integer $j$ this symmetry relates operators to
shadow operators in the conformal field theory. 
The value of $j$ at
this minimum, $j_0(\lambda)$,  corresponds to the location of the
conformal Regge intercept associated with a given spectral curve
$\Delta(j)$.   In  strong coupling, the Pomeron intercept~\cite{Brower:2006ea,Kotikov5}, 
\be
\alpha_P=j_{0}(\lambda)=2 - \frac{2 }{\lambda^{1/2}}+ O(1/\lambda)\, ,
\label{eq:P-intercept}
\ee
was determined by perturbing the spectral curve $\Delta_P(j)$ about the graviton at $j=2$ in the super-gravity limit.
 The same principle can be applied to  weak coupling, with the minimum identified with the  BFKL Pomeron
intercept~\cite{Lipatov:1976zz,Kuraev:1977fs,BL,Lipatov:1985uk,KirschLipat}
\be
\alpha_P=j_{0}(\lambda)=1 +  \frac{4 \ln 2 }{\pi} \lambda + O(\lambda^2) \,.
\label{eq:P-intercept_BFKL}
\ee 
A major challenge in  ${\cal N}=4$ SYM is to
determine this intercept $j_0(\lambda)$  for
all $\lambda$ in the  large $N_c$ approximation and to apply
this analysis to other trajectories \cite{Alfimov:2014bwa}.

In this paper we apply an expansion procedure for  spectral curves, that describes short strings in AdS \cite{Gromov:2011de,Gromov:2011bz,Basso:2011rs}, to improve the strong coupling expansion for both the leading conformal Pomeron (C =+1) and  Odderon (C= -1) trajectories.   For the Pomeron this is
a simple extension of earlier results~\cite{Costa:2012cb, Kotikov:2013xu,Janik:2013pxa,Gromov:2014bva}, while the application to the Odderon trajectory~\cite{Brower:2009,Brower:2013jga} is new. 
For   ${\cal N}=4$ SYM, the C= +1  exchange  is associated with
exchange of local operators with $\pm$  light-cone  components, 
\be
 {\cal O}^\pm_P(j,k)=Tr [F^{\pm \perp}(D^{\pm})^{j-2} F_\perp^\pm Z^k] +\cdots\,, 
 \label{eq:O_P}
\ee
 with $ j=2,4,6,\cdots$ and   $k=0,1,\cdots$. Here,  $Z$ is a scalar field with $SU(4)$ R-charge.   The leading Pomeron  has  $k=0$, but we extend known results to the case of R-charge exchange with $k\neq 0$. Due to super-symmetry, the analysis can be simplified by relating the relevant spectral curves  to that  for single-trace operators in the $sl(2)$
sector, symbolically expressed as $Tr [(D^{\pm})^{j-2}Z^{k+2}]+\cdots$~\cite{D'Hoker:2002aw,Kotikov:2007cy}. For the case of the $C=-1$ Odderon trajectory, we shall show how similar techniques can also be applied~\footnote{Preliminary result was first reported at  the ``Low-x Workshop", 2013, Rehovot and Eilat, Israel \cite{Brower:2013jga}.}.  

\begin{figure}[t]
\begin{center}
\includegraphics[width = 2in]{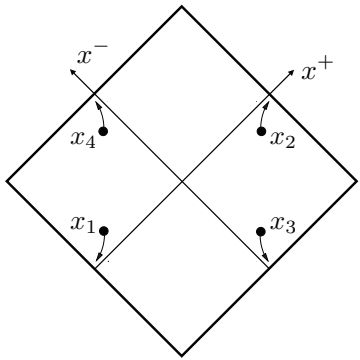}
\caption{Conformal compactification of the  light-cone showing points taken
  to null infinity in the Regge limit.
In light-cone coordinates $(x^+,x^-,x_\perp)$ we take  $- x^+_1 \sim x^+_2\to\infty$  and $- x^-_3 \sim x^-_4\to\infty$,
keeping $x^i_\perp$ fixed.}
\label{fig:ReggeLimit}
\end{center}
\end{figure}

We shall restrict our attention to the pure conformal limit and begin here by briefly describing the  Regge limit in the context of conformal field theories \cite{Cornalba:2007fs,Cornalba2008,Cornalba:2009ax,Costa:2012cb}.  Consider the connected component for a 
four point correlation function of primary operators ${\cal O}_i$ of dimension $\Delta_i$. 
Defining  $x_{ij}=x_i-x_j$, we have
\be
A(x_i) = \langle {\cal O}_1(x_1) {\cal O}_2(x_2){\cal O}_3(x_3){\cal O}_4(x_4)\rangle_c =
  \frac{1}{(x^2_{12})^{\Delta_1}(x^2_{34})^{\Delta_3}} 
  \,F(u,v)\,, \label{eq:A(X)}
\ee
where  
\be
u=\frac{x_{12}^2x_{34}^2}{x_{13}^2x_{24}^2}\,, \quad\quad
v=\frac{x_{14}^2x_{23}^2}{x_{13}^2x_{24}^2} \,,
\ee
are the cross ratios  and for simplicity we have assumed
$\Delta_1=\Delta_2$ and $\Delta_3=\Delta_4$.
We need to examine  the double light-cone  limit of  vanishing $x^2_{12}$ and $x_{34}^2$, which corresponds to $u\rightarrow 0$ and $v\rightarrow 1$ in a Minkowski setting. From the perspective of light-cone  OPE, this  limit can be reached  by scaling 
$x_1^+ \to \lambda x_1^+$, $x_2^+ \to \lambda x_2^+$, $x_3^- \to \lambda x_3^-$, $x_4^- \to \lambda x_4^-$
with $\lambda\to\infty$, keeping the causal relations $x_{14}^2,
x_{23}^2<0$, as illustrated in Fig. \ref{fig:ReggeLimit}. 
In a frame where $x_{1\perp}=x_{2\perp}$ and $x_{3\perp}=x_{4\perp}$,   this corresponds to approaching the respective null infinity
while keeping the relative impact parameter
\be
b_\perp= x_{1\perp}-x_{3\perp}\,,
\label{eq:Impact}
\ee
fixed.
In terms of the cross ratios, this corresponds to $u \rightarrow 0$ and $v\rightarrow 1$, with   $(1 - v)/\sqrt{u}$ fixed. 
Alternatively, defining $u = z\bar z$ and $v = (1 - z) (1 - \bar z)$
with $z =\sigma e^{\rho}$ and $\bar z = \sigma e^{-\rho}$, the precise Regge limit
can be specified by    $\sigma \rightarrow 0$ for fixed $\rho$.   

Using Regge theory the leading behavior of the reduced correlation functions $F$ can be determined by the intercept $j_0$, computed from
the corresponding leading spectral curve $\Delta(j)$ described earlier, with a general form
\be
F(\sigma,\rho) \approx  f(\rho) \, \frac{\sigma^{1-j_0}}{|\ln\sigma|^{3/2}}\,.
\label{eq:reducedF}
\ee
This is entirely analogous to conventional Regge theory where the spectrum of Regge trajectories determines high energy scattering amplitudes.

The paper is organized as follows. In Sec. \ref{sec:OPE},  we  will review the dictionary that translates the
Regge description from the language of CFT and OPE into scattering amplitudes in AdS space. This will allow us
to proceed directly within the framework of CFT's without explicit recourse to the AdS/CFT correspondence. Nonetheless
reference to AdS space can and will be made to provide additional intuition. 
In Sec. \ref{sec:Pomeron}, we analyze the Pomeron intercept beyond the
``diffusion limit", (\ref{eq:P-intercept}), leading to 
a systematic expansion for the Pomeron intercept in $\lambda^{-1/2}$.   A similar analysis is done for Odderons in Sec. \ref{sec:Odderon}.  
In particular, we clarify how one of the Odderon solutions has
intercept that remains fixed at 1 to all orders in $1/\sqrt\lambda$. 
These results are summarized and discussed in
Sec. \ref{sec:Discussion} where we also clarify further the 
all-coupling formula proposed by Basso  in \cite{Basso:2011rs}, and
its possible generalization. We also provide additional comments relating to the weak coupling limit and other  issues.

\section{Conformal Regge Representation}
\label{sec:OPE}

While a conformal Regge analysis can be presented entirely in a CFT  language
\cite{Cornalba:2007fs,Cornalba2008,Cornalba:2009ax,Costa:2012cb}, it is often useful to
follow  the earlier derivations in invoking  a scattering process in $AdS$~\cite{Brower:2006ea,Brower:2007qh,Brower:2007xg,Cornalba:2006xm,Cor}.  
Although  both approaches are equivalent,  they  offer separate intuitive frameworks.   

\noindent{\bf Regge Theory from  CFT Partial Wave Expansion:} The t-channel OPE conformal partial-wave expansion for the connected component of a 4-point function  is given 
by a sum over conformal blocks,
\be
F(u,v)= \sum_{j}  \sum_\alpha C^{(12),(34)}_{\alpha,j}
\,  G( j,\Delta_{\alpha}(j);u,v) \, . 
 \label{eq:ConformalBlock}
\ee
For the Regge limit (on the light cone), these are Minkowski conformal blocks, 
defined with appropriate boundary conditions,
or equivalently, as analytic continuation from Euclidean space
as explained in
Refs.~\cite{Cornalba:2007fs,Cornalba2008,Cornalba:2006xk}.  Although, for planar
${\cal N}=4$ SYM  in the Regge limit, we shall restrict the 
sum to  single-trace conformal  primary operators, a completely
general representation can be found by introducing basis function for
 the  principle unitary conformal  representation and expanding the  amplitude as
\be
F(u,v)= \sum_{j} \int^{\infty}_{-\infty} \,\frac{d \nu}{2\pi } \, a  (j, \nu) \, {\cal G} (j,\nu ; u,v)\,.
\label{eq:groupexpansion}
\ee 
This group representation combines a discrete sum in the spin $j$ and a  Mellin transform in  a complex $\Delta$-plane,
with $\Delta = 2 + i\nu$, as explained by Mack  \cite{Mack:2009mi} and others~{\footnote{We have absorbed factors coming from Plancherel measure, etc., into the partial-wave amplitude $a  (j, \nu) $ and will also normalize the conformal harmonics, ${\cal G} (j,\nu ; u,v)$, so it eventually leads to conformal blocks with conventional normalization. It has also been demonstrated in \cite{Mack:2009mi} how the CFT ``Mellin-representation" can be expressed  in this group-theoretic form.}. The conformal harmonics, ${\cal G} (j,\nu ; u,v)$, are eigen-functions  of the quadratic Casimir operator of $SO(4,2)$. To recover the standard conformal block expansion, it  is conventional to close   the contour in the lower-half $\nu$-plane~\footnote{Due to conformal invariance,  the integrand  is even in $\nu$, or, equivalently, symmetric in $\Delta\leftrightarrow 4-\Delta$. The contour can be closed either in the upper or the lower  half $\nu$-plane.  The poles in the upper half $\nu$-plane corresponds to ``shadow" operators.},
(equivalently, closing the contour in $\Delta$-plane  to the right), picking up only
dynamical poles in $a(j, \nu)$, at $\nu(j)=-i( \Delta(j)-2)$. 
 After summing over these pole
contributions, one arrives at the conformal partial wave expansion 
(\ref{eq:ConformalBlock}), which also serves as an OPE.    These
dynamical poles correspond to all allowed
conformal primaries, ${\cal O}_{\Delta(j)}$, with spin $j$ and
dimension $\Delta(j)$.

We are now faced with a discrete sum over spin $j$. A distinguishing feature of  the Regge limit is the fact that  the
conformal harmonics,
\be
{\cal G} (j,\nu ; u,v) \sim \sigma^{1-j}\Omega_{i\nu}(\rho) \, , \quad\quad
\Omega_{i\nu}(\rho)= \frac{1}{4\pi^2}\, \frac{\nu \sin (\nu \rho)
}{\sinh \rho}\, ,
\label{eq:PartialWave}
\ee
 are more and more divergent  for  increasing  $j>1$ as  $\sigma\rightarrow 0$. 
Therefore one cannot take the Regge limit term by term in
(\ref{eq:groupexpansion}}).  The traditional Regge hypothesis is that
this sum can be evaluated by representing the partial wave
expansion by the Sommerfeld-Watson transform in the analytic j-plane.  For
conformal Regge theory,  this step leads to a double-Mellin representation \cite{Cornalba:2007fs,Cornalba2008,Cornalba:2009ax},
\be
F(u,v)= - \int_{-i\infty}^{i\infty}  \frac{dj}{2\pi i}   \;      \frac
{1 \pm e^{-i \pi j } }{\sin\pi j }   \int^{\infty}_{-\infty}
\frac{d\nu}{2\pi }  \,\, a(j, \nu)\,\,  {\cal G} (j,\nu ; u,v) \, ,
\label{eq:newgroupexpansion}
\ee
where the contour in $j$ is  to stay to the right of singularities of $a(j, \nu)$.  
Note that in (\ref{eq:newgroupexpansion}) we must consider 
separate expressions for  even or odd spin values, which will
correspond to $C=\pm 1$ contributions respectively. 
While a direct proof of this $ j$-plane representation is lacking
for CFT's in general, it has been shown to hold at
strong coupling on the basis of the AdS/CFT correspondence for ${\cal N}=4$ SYM \cite{Brower:2006ea} and 
at weak coupling in the BFKL limit
\cite{Cornalba:2007fs,Cornalba2008}. Moreover
it is a natural assumption in order that non-conformal deformation
give back the traditional Regge representation. 
This double-Mellin representation for conformal Regge theory
leads to a meromorphic representation in the $\nu^2-j$ plane, with
poles specified by the collection of allowed spectral curves,
$\Delta_\alpha(j)$.  Still we should emphasize that this  conformal ``Regge pole hypothesis'' 
is similar but is neither identical or a consequence of the
conventional Regge theory. The
conventional Regge $j$-plane analyticity with moving singularities in the
$j-t$ plane, is replaced by  analyticity and moving singularities in this
$j-\nu^2$ plane. In Ref.~\cite{Brower:2006ea} 
this distinction is clearly
delineated by introducing a confining deformations of the Poincare
patch of  AdS space which interpolate  smoothly between conformal and
non-conformal Regge theory.

As an illustration, let us re-consider the  $C=+1$ case and focus on
the contribution from a single conformal Pomeron pole in $\nu^2$,
\be
a(j,\nu) = \frac{r(j)}{\nu^2 + (\Delta(j) -2)^2 }\,,
\label{eq:partial-wave}
\ee
 characterized by a spectral curve $\Delta(j)$, with the residue $r(j)$ which vanishes  at $j =  0, -2, -4, \cdots$. 
  Closing the contour first   in the lower-half $\nu$-plane,  (\ref{eq:newgroupexpansion}) leads  to a single-Mellin representation
\be
F(u,v) \approx - \int
\frac{dj}{2\pi i}   \;    \frac {1 + e^{-i \pi j } }{\sin\pi j } \, r(j) \,\, \sigma^{1-j} \,\,\frac{e^{(2-\Delta(j)) \rho}}{\sinh \rho}\,,
 \label{eq:Single-Mellin}
\ee
where we have taken the limit of $\sqrt u=\sigma$ small,  with  $(1-v)/\sqrt u\approx 2 \cosh \rho$ fixed.  In this limit, the dominant contribution comes from the right-most singularity in the $j$-plane, which enters through $\Delta(j)$.
Consider next  the spectral curve $\Delta(j)$ for Pomeron exchange and focus on the strong coupling limit
where one has    $\Delta(j) = 2+\sqrt  2\lambda^{1/4} \sqrt {j -j_0}$,
 with $j_0$ given by (\ref{eq:P-intercept}).  Observe that  this gives a fixed branch-cut in the $j$-plane.  In the 
 limit $\sigma\rightarrow 0$, by pushing the $j$-contour to the left and picking up the contribution from the branch-cut at $j=j_0$, (\ref{eq:Single-Mellin})  leads directly to the singular behavior (\ref{eq:reducedF}), as promised.  A similar result is also obtained if one considers the
 weak coupling expansion of the spectral curve $\Delta(j)$, but now $j_0$ in  (\ref{eq:reducedF}) is given by the BFKL expansion
 (\ref{eq:P-intercept_BFKL}).

\noindent{\bf AdS Impact parameter representation:} Now let us turn  to a momentum space treatment for the Regge limit in CFT. 
Consider the Fourier transform of the connected correlation function   defined in (\ref{eq:A(X)}),
\be
(2\pi)^4\, \delta^{(4)}\!\left( \sum p_j \right) i\, T(p_1, p_2, p_3, p_4) =  
\left\langle  {\cal O}_1(p_1) {\cal O}_2(p_2) {\cal O}_3(p_3) {\cal O}_4(p_4) \right\rangle_c\,.
\ee 
The  amplitude $T(p_j)$  can be expressed as a function of Mandelstam
invariants $s$, $t$, and $p_j^2$. The  Regge limit corresponds to $s$ large, which defines a light-cone  direction, with $t<0$ and $p_j^2$ fixed. In this limit,  the momentum transfer is asymptotically transverse, with $t=(p_1+p_2)^2\approx - q_\perp^2$. 
Using conformal symmetry,  it is possible to express the amplitude $T(p_j)$ in an $AdS$ impact parameter representation, which in the 
Regge limit takes the form~\cite{Cornalba:2006xm,Cor,Cornalba:2007fs,Cornalba2008}
\bea
T(s,t,p^2_i) & \approx &   \int \frac{dz}{z^5}\, \frac{d{z'}}{{z'}^5} \,
\Phi_1(z,p_1^2)\Phi_{2}(z,p^2_2) \,  { \cal G}(s,t,z,z')   \, \Phi_{3}(z',p_3^2)\Phi_{4}(z',p_4^2)
\; , \nn
 { \cal G}(s, t,z,z') &= & (zz')^2s\,  \int \frac{d^2b_\perp}{4 \pi^2}\, e^{\textstyle i q_\perp \cdot b_\perp}  {\cal T}(S,L) \; , \label{IPrep}
\eea
with $b_\perp$ the two-dimensional impact parameter introduced earlier (\ref{eq:Impact}).  
The amplitude ${\cal T}={\cal T}(S, L)$  encodes all dynamical information and, due to conformal symmetry, depends only on 
the variables
\begin{equation}
S=zz' s \ ,\ \ \ \ \ \ \ \ \ \ 
\cosh L=\frac{z^2+z'^2+b_\perp^2}{2zz'}\,.
\label{S,L}
\end{equation}
The same representation was  obtained  through  direct  AdS/CFT considerations
~\cite{Brower:2006ea,Brower:2007qh,Brower:2007xg},  via   generalized Witten diagrams, string vertex operators, etc., 
leading to a Regge kernel, ${\cal K}( s, b_\perp, z,z')$. Up to irrelevant constants, this kernel is related to the amplitude ${\cal T}(S,L)$ by~\footnote{ The   $1/N^2$ dependence in  ${\cal T}(S,L )$, expected for $AdS$
gravitational interactions,  is normally removed from ${\cal K}$. In \cite{Brower:2007qh,Brower:2007xg}, a reduced variable $\widetilde s=S=zz's$ was used extensively. One occasionally also used $\eta$ or $\xi$, instead of $L$.}
\be
{\cal K}(s,b_\perp,z,z') \sim  {N^2}\,{(zz')^2\, s}\,\,{\cal T}(S,L ) \,.
\ee
The Regge limit is now $S\to \infty$ with fixed $L$.
It is important to note that the conformal  representation (\ref{IPrep}) of the amplitude is valid for any value of the coupling constant, since it relies
only on conformal invariance. 
However, it is quite natural from the view point of the dual $AdS$ scattering process, where
transverse space is precisely a three-dimensional hyperbolic space $H_3$, whose boundary is conformal to the physical transverse 
space $\mathbb{R}^2$. The cross ratio $L$ is then  identified with the geodesic distance between two points in $H_3$ 
that are separated by $b_\perp$ 
along $\mathbb{R}^2$ and have radial coordinates $z$ and $z'$.
The other cross ratio $S=zz's$ measures the local energy squared of the scattering process in $AdS$,
since $z$ and $z'$ define the local $AdS$ scales for each incoming particle.
Moreover, the functions $\Phi_i(z,p_i^2)$ are $AdS$ bulk to boundary propagators with a plane wave source of momentum $p_i$ 
created by the gauge theory operator at the boundary $z\to 0$. 
To define on-shell scattering for non-conformal  amplitudes, $T(s,t)$,
as introduced in Ref.~\cite{Brower:2006ea},
one  deforms the dual AdS space in the IR, breaking conformal
symmetry,  and replaces  $\Phi_{i}(z,p^2_i)$ by normalizable  wave functions for 
``hadronic'' (or glueball) eigenstates~\footnote{
Note that at weak coupling the product of wave functions $\Phi_1\Phi_2$ and  $\Phi_3\Phi_4$ are similarly
replaced by  the dipole parton distributions for the external
particles~\cite{Cornalba2008}, so  that this  impact parameter
representation (\ref{IPrep}) is maintained.}.

By considering  the radial Fourier decomposition in the $AdS$ impact parameter space $H_3$, one can  derive  a double Mellin representation
of the kernel, as done earlier for CFT analysis in a coordinate representation, Eq. (\ref{eq:newgroupexpansion}).  
This radial decomposition simply considers harmonic functions $\Omega_{i\nu}$ on  $H_3$, given by (\ref{eq:PartialWave}), which satisfy
$(\nabla^2+1+\nu^2)\,\Omega_{i\nu}=0$, where $\nabla^2$ is the $H_3$
Laplacian. This is equivalent  to introducing  ${\mathbb R}^2$ harmonic functions, $e^{iq_\perp b_\perp}$, in the standard impact parameter decomposition. The only difference is that now we have a scattering process in $AdS$ space.
From a CFT point of view, this representation can also be derived by writing the conformal partial wave decomposition 
of the amplitude ${\cal T}(S,L)$, and then taking the Regge limit.
Thus, as before, the $\nu$-integral reflects conformal invariance due to dilatation, and the $j$-integral represents a coherent sum of $t$-channel spin fields, as was done earlier for the  OPE sum via a Sommerfeld-Watson transform.
 To be more explicit,  since   the  Pomeron/Odderon kernels receive contributions respectively from all
even/odd spins,   these kernels can be expressed as 
\be
{\cal K}_\pm (s, b^2_\perp, z,z') =  - (zz')\int \frac{dj}{2\pi i } \frac{1 \pm e^{-i \pi j}}{\sin\pi j}  \int_{-\infty}^{\infty} \frac{d\nu}{2\pi} \,S^{j} \,\,  G_\pm(j,\nu)\, \Omega_{i\nu}(L)\,.
  \label{MellinJ}
\ee
 This representation is a  consequence of conformal invariance, which must next be supplemented by dynamics, i.e., specifying the Pomeron/Odderon propagator $G(j,\nu)$.
 
By examining Witten diagrams for  exchanging  
spin-$j$ fields  in the Regge limit and also their string duals,  one
is led to $G(j,\nu)$ having a simple pole in the $\nu^2$-plane,
determined by the spectral curve  $\Delta(j)$ associated with these
fields,  exactly as in coordinate treatment (\ref{eq:partial-wave}).
Here $G(j,\nu)$ can be  related  to the $\nu$-transform of a transverse scalar bulk-bulk propagator with an effective $j$-dependent $AdS$ mass.   
The residue at this pole can be  related to the local $AdS$ coupling of  the exchanged fields to the external states. 
Upon closing the $\nu$-contour, one picks up a factor
$
G(j,b_\perp, z,z')\sim  {e^{ -(\Delta(j) - 2 )L }}/{  \sinh L}\, .  \label{eq:Gpx}
$
Finally, for  Pomeron(Odderon) exchange,  by identifying the $j$-plane  branch-point at $j_0$ associated with the Pomeron(Odderon) spectral curve, from (\ref{MellinJ}) one has  for $S$ large, by pulling the contour to the left,
\be
{\cal K}(s,b^2_\perp, z,z')  \approx  (zz')   \tilde{f}(L) \, \frac{S^{j_0}}{| \ln S |^{3/2}}\,,
\label{eq:final-K}
\ee
just as in the coordinate representation \eqref{eq:reducedF}.

\noindent{\bf Regge Dictionary for CFT:} We have therefore two
representations of the correlation function in the Regge limit. One
derived from the CFT analysis in position space $F(u,v)$, given by
(\ref{eq:newgroupexpansion}), and another from a computation in
momentum space with a clear geometrical interpretation as a scattering
process in $AdS$, given by (\ref{MellinJ}). This establishes a
dictionary, where, in the Regge limit, 
\bea 
F(u,v)&\leftrightarrow&{\cal T}(S,L)=N^{-2}\, (zz' )^{-2} s^{-1} {\cal K}( s,b^2_\perp, z,z')\,, \nn 
\sigma =\sqrt u &\leftrightarrow& S^{-1}= (zz' s)^{-1}\,, \nn 
\cosh \rho \approx \frac{1-v}{2\sqrt u} &\leftrightarrow& \cosh L = \frac{b_\perp^2 + z^2 + {z'}^2}{2 z z'}\,.  \label{eq:CFT-Regge-dictionary}
\eea 
Although it is possible to carry out a more formal analysis in
establishing this equivalence, we will not pursue this here \cite{Cornalba:2009ax}. It
suffices to emphasize the exact equivalence of the two approaches  to identify
the spectral curve, $\Delta(j)$ in Fig. \ref{fig:BFKLDGLAP},  which serves as the common link between them.

 \section{Conformal Pomeron}
 \label{sec:Pomeron}

 The Pomeron spectral curve  $\Delta_P(j)$ in the strong coupling limit, Fig. \ref{fig:BFKLDGLAP}, can be obtained by  an intuitive derivation  based on a  flat-space leading closed-string linear trajectory. Through AdS/CFT, this simple result can   be understood as a perturbation about  the traceless-transverse graviton mode,  $\nabla^2 h_{MN}=0$, with $j=2$ and $\Delta=4$ in the $\lambda=\infty$  super-gravity limit.  Here 
 $\nabla^2$ is the tensor Laplacian on $AdS_5$. 
Let us now consider the limit of $j\to 2$ and $\lambda\to\infty$ with  $\sqrt{\lambda}(j-2)$ fixed.
This limit can be understood by introducing a Reggeon vertex operator, ${\cal V}^\pm$, on the
string world sheet in a weakly curved target $AdS_5\times S^5$ space~\cite{Brower:2006ea}.
This Reggeon vertex operator depends on  $(j,\nu, t)$  for the  $O(4,2)$ Casimir, and on $k=\tau-2$ in case
we wish to consider exchange of $SO(6)$ $R$-charge.
The effect of the Reggeon operator is to  resum the  exchange  of all modes in the leading Regge trajectory 
with even positive integral spins, which leads to the effective Regge spin
\be
j = 2 -  \frac{\tau^2+\nu^2}{2 \sqrt \lambda} \, .
\label{eq:diffusion}
\ee
Then the world-sheet Virasoro on-shell condition, $L_0{\cal V}^\pm=\bar L_0{\cal V}^\pm={\cal V}^\pm$,
establishes the relation 
\be
\Delta_P(j,\tau) = 2+ i\nu  \label{eq:BPST2}
\ee
where $\Delta_P(j,\tau)$ is the continuation of the anomalous dimension curve for the exchanged 
gauge theory operators in the leading Regge trajectory ${\cal O}^\pm_P(j,\tau)$ given in (\ref{eq:O_P}).
By following this procedure, one relates the exchange of $AdS$ higher spin fields to the 
dual gauge theory operators with spectral curve $\Delta_P(j,\tau)$. 

In the above double limit the Regge spin has a quadratic dependence in the dimension $\nu$, also
known as the diffusion limit (in a weak coupling expansion we may also consider such a diffusion limit). 
The terminology stems from the fact that the kernel in momentum space  takes on a diffusion form at $t=0$,
with diffusion time $\ln (zz's)$ fixed by the $AdS$ energy $S=zz's$~\cite{Brower:2006ea}.
In this limit we can already observe that the spectral curve $\Delta_P(j,\tau)$ has a branch point at $j_0$, 
\be
\Delta_P(j) = 2 +  \sqrt{2\lambda^{1/2}} \sqrt {j-j_0}\, .   
\label{eq:singularity}
\ee
where 
\be
j_0=2-\frac{\tau^2}{2\sqrt{\lambda}}\,,
\ee
which generalizes (\ref{eq:P-intercept}) for the case of twist $\tau$.  

Beyond the diffusion limit (\ref{eq:diffusion}) the Reggeon spin $j=j(\nu,\tau)$ admits the strong coupling expansion 
\be
j(\nu,\tau) = 2 -  \frac{\tau^2+\nu^2}{2 \sqrt \lambda}  \left( 1 + \sum_{n=2}^\infty \frac{\tilde{j}_n(\nu^2,\tau)}{\lambda^{(n-1)/2}}  \right)  \, ,
\label{ReggeonSpin}
\ee
which is a simple generalisation to  arbitrary twist $\tau$ of the results presented in \cite{Cornalba:2007fs,Costa:2012cb}
(such that at infinite coupling, for $j=2$, the dimension of the operator is given by the protected value of $2+\tau$).
Notice that $j(\nu,\tau)$ must be an even function of $\nu$ to implement the symmetry property 
$\Delta_P(j,\tau)\leftrightarrow 4-\Delta_P(j,\tau)$.
The function $\tilde{j}_n(\nu^2,\tau)$, defined for $n\ge 2$, is a polynomial of degree $n-2$,
\be
\tilde{j}_n(\nu^2,\tau)=\sum_{k=0}^{n-2} c_{n,k}\nu^{2k}\,,
\ee
with $\tau$-dependent coefficients $c_{n,k}$.
This follows from the requirement that the $AdS$ amplitude has a well defined flat space limit \cite{Cornalba:2007fs}.
Consistency with the strong coupling expansion of the spectral curve $\Delta_P(j,\tau)$ further restricts this
polynomials to have smaller degree~\cite{Costa:2012cb}, more precisely, for $n\ge4$
\be
c_{n,k}=0\ \ {\rm for}\ \ \left[ \frac{n}{2} \right] \le k\le n-2\,,
\ee
as also confirmed in \cite{Janik:2013pxa}.

Eq. (\ref{ReggeonSpin}) corresponds to an expansion for  the Reggeon spin about the symmetry point $\Delta=2$, ($\nu^2=0$), in the strong coupling limit, subject to the constraint that $j=2$ at $\nu^2=-\tau^2$. In the next subsection we review recent results for the strong coupling expansion of the spectral curve $\Delta_P(j,\tau)$ about $j=2$ that will
allow us to compute the pomeron spin  $j(\nu)$, i.e., the inverse of the spectral curve $\Delta(j)$, with  $\Delta$ and $\nu$ related by (\ref{eq:BPST2}), beyond the diffusion limit at arbitrary twist $\tau$.
The discussion leads to a unified picture that can also be applied to the Odderon Regge trajectories, in Sec. \ref{sec:Odderon}.

\subsection{Pomeron spin versus anomalous dimensions at strong coupling}

Much attention has been paid in recent years to the study of anomalous
dimensions for composite operators of ${\cal N}=4$ SYM.  
Because of supersymmetry, many related
operators share the same anomalous dimensions.  It is generally
believed that, due to integrability \cite{Minahan:2002ve}, scaling dimensions for gauge
invariant operators can be efficiently calculated for all 't Hooft
coupling,  in the large-$N$ planar limit,  via the
so-called TBA/Y-system and its generalizations \cite{Beisert:2010jr}.
These operators and their cousins can be treated as generalized
Heisenberg spin chains.  For instance, the weak-coupling one-loop anomalous
dimension $\gamma(S,\tau)$ of single-trace operators in the $sl(2)$
sector, symbolically expressed as $Tr [D_\pm^{S} Z^\tau]+\cdots$, can be
calculated explicitly,  with   $\Delta(S,\tau) = S + \tau +\gamma(S,\tau)$
 their scaling dimension.  However, beyond
one-loop, and particularly for short operators ($S$ and $\tau$ small), analytic solutions have been difficult to obtain.
For strong coupling, conformal  dimensions can  be calculated
semi-classically in a world-sheet sigma model approach around soliton
solutions~\cite{Gubser:2002tv,Frolov:2002av,Arutyunov:2003uj}, as semi-classical treatment for GKP strings. This leads to a strong coupling loop-expansion, with $1/\sqrt \lambda$ playing the role of $\hbar$, 
\be
\Delta
 = \lambda^{1/4}\left(\delta_{0}+\frac{\delta_{1}}{\lambda^{1/2}}+\frac{\delta_{2}}{\lambda}+
 \frac{\delta_{3}}{\lambda^{3/2}}+\frac{\delta_{4}}{\lambda^{2}}+\cdots\right) \, ,\label{eq:loop}
\ee
with   $S$ and $\tau$  dependence entering the  $\ell$-loop contribution $\delta_{\ell}$  through a scaling hypothesis~\footnote{$\delta_{\ell}$   is  assumed to be a function of  ${\cal S}=S/\sqrt\lambda$ and ${\cal T}=\tau/\sqrt \lambda$.  Occasional discussions for   small $S$ and/or  small $\tau$  are typically based on extrapolation under this scaling hypothesis. As such,  less attention has been paid in the past to the symmetry property in $\Delta$. For an alternative but   related study,  see~\cite{Vallilo:2011fj,Mazzucato:2011jt}.}. 
However, calculation beyond 1-loop is impractical. In most approaches of this type, the emphasis has been on long strings.

One analysis of particular interest to us is the expansion for the
spectral curve about the point $S=0$, 
\be
\label{eq:Basso_expansion}
\Delta_Z(S,\tau) = \tau + \alpha_1(\tau,\lambda) S +
\alpha_2(\tau,\lambda) S^2 + \alpha_3(\tau,\lambda) S^3+\cdots \ee
where $\Delta_Z(0,\tau)=\tau$, since the operator is 1/2-BPS and its
dimension is protected. The  form of the ``slope function",
$\alpha_1(\lambda,\tau)$, has recently been conjectured by Basso for all 't
Hooft coupling~\cite{Basso:2011rs} (see also \cite{Basso:2012ex,Gromov:2012eg}), which can be expressed in a compact form in terms of
Bessel functions, 
\be
 \alpha_1(\lambda,\tau) =
\frac{\sqrt\lambda}{\tau}\, Y_\tau(\sqrt\lambda) \,, \label{eq:Basso.a1} 
\ee
{\it for all $\lambda$}, with $Y_\tau(x)=I'_\tau(x)/I_\tau(x)$, where $I_\tau(x)$ is the $\tau$-th modified bessel function. 
At  weak coupling, $\alpha_1(\lambda,\tau)=1+O(\lambda)$,  
and at the strong coupling, $\alpha_1(\lambda,\tau)=\sqrt\lambda/\tau+O(1/\sqrt \lambda)$.
This result was first derived for $\tau=2$ as a solution to the
``asymptotic Bethe ansatz" (ABA) equations. It has been argued in
\cite{Basso:2011rs}, with further support in
\cite{Basso:2012ex,Gromov:2012eg}, that this holds for all $\tau>2$,
for the configurations with ``minimum mode numbers"~\footnote{We will
  return to a discussion on this and related issues in
  Secs. \ref{sec:Odderon} and \ref{sec:Discussion}.}.  More recently,
the second coefficient, $\alpha_2(\lambda,\tau)$, has also been
calculated numerically, but it is not possible at this moment to
express it in a closed form in terms of elementary
functions~\cite{Gromov:2014bva,Beccaria:2014rca}.
Due to super-symmetry, it is known that the Pomeron spectral curve~\footnote{In
  order to avoid notational confusion, in what follows, instead of
  $S$, we shall switch to $j = S+2$, e.g., for the $sl(2)$ sector, we
  have $Tr [D_\pm^{j-2} Z^\tau]$, instead of $Tr [D_\pm^S Z^\tau]$. } is
directly related to $\Delta_Z(S,\tau)$ at $\tau=2$ by~\cite{D'Hoker:2002aw,Kotikov:2007cy} 
\be 
\Delta_P(j) = 2+ \Delta_Z(j-2,2)\,. \label{eq:BPST.2} 
\ee 
Therefore, these recent analyses,
appropriately generalized, can be applied to our study of conformal
Pomeron and Odderon, particularly in the large $\lambda$ limit.

For planar ${\cal N}=4$ SYM, it is possible to generalize our discussion for the Pomeron spectral curve to include CFT operators  with $R$ charges, which through AdS/CFT, amounts to allow fluctuations in $S^5$. 
For our purpose, as already described in (\ref{eq:O_P}), the  relevant CFT operators 
are
\be
{\cal O}_P(j,\tau)=Tr\left[ F_{\mu\sigma} D_{\rho_1}\cdots D_{\rho_s} F_{\sigma\nu} Z^{\tau-2}\right] + \cdots,  \label{eq:Pomeron-S5}
\ee
with  $\tau\equiv 2+k \geq 2$.
For the leading Regge singularity, we will be dealing with  the
light-cone components $Tr\big[ F_{\pm\perp} D_{\pm}\cdots D_{\pm} F_{\perp\pm} Z^{\tau-2}\big] + \cdots$.     
The generalized Pomeron spectral curve  $ \Delta_P(j,\tau)$ can be expanded using \eqref{eq:BPST.2} and \eqref{eq:Basso_expansion}  around $j=2$, leading to
\be
 \Delta_P(j,\tau)= 2+\tau + \alpha_1(\lambda,\tau)(j-2) + \alpha_2(\lambda,\tau) (j-2)^2+ \alpha_3(\lambda,\tau) (j-2)^3 +\cdots\,.  
 \label{eq:Basso.a2}
\ee
where (\ref{eq:Basso.a1})  applies. 

Because of the symmetry under $\Delta_P(j,\tau)\leftrightarrow
4-\Delta_P(j,\tau)$, we again require the function $\Delta_P(j,\tau)$ to have a square-root singularity at $j_0(\tau)$, with $j_0(\tau)=2-O(\lambda^{-1/2})$. This branch point renders  the expansion (\ref{eq:Basso.a2}) with  a radius of convergence which vanishes as $\lambda^{-1/2}$, leading to  expansion coefficients which grow as $\alpha_n\sim O(\lambda^{n/2})$. Nevertheless, a convergent expansion can be achieved by considering the symmetric combination $\big(\Delta_P(j,\tau) - 2\big)^2=-\nu^2$, for which this square-root branch point is absent. This in turn leads to a convergent  expansion in the strong coupling limit, 
 \be
\big(\Delta_P(j,\tau\big) - 2)^2 = \tau^2 + \beta_1 (\lambda,\tau)(j-2) + \beta_2(\lambda,\tau) (j-2)^2 + \beta_3(\lambda,\tau) (j-2)^3+ \cdots\,,  \label{eq:Basso.beta}
\ee
where  $ \beta_1 (\lambda,\tau)=2\tau \alpha_1(\lambda,\tau)$, $\beta_2 (\lambda,\tau)= \alpha_1(\lambda,\tau)^2 + 2\tau \alpha_2(\lambda,\tau)$, etc.  
Based partly on semi-classical analysis of GKP strings~\cite{Gubser:2002tv,Frolov:2002av,Arutyunov:2003uj}, 
one  expects a radius of convergence of the order $O(\lambda^{1/2})$. Consistency with (\ref{eq:singularity}) and the existence of a smooth super-gravity limit then require that $\beta_n\sim O(\lambda^{(2-n)/2} )$, so that    each coefficient $\beta_n(\lambda)$ in turn admits an expansion, 
\be
\beta_n = 2\lambda^{\frac{2-n}{2}}\Big( b_{n,0} + \frac{b_{n,1}}{\lambda^{1/2}}+ \frac{b_{n,2}}{\lambda}+ \frac{b_{n,3}}{\lambda^{3/2}}+\cdots\Big)\,,\label{eq:Basso.bn}
\ee
where we have taken the factor of 2 out so that later on the expansion will be normalized with $b_{1,0} = 1$, and the 
coefficients $b_{n,m}$ are in general $\tau$-dependent.
 The viability of the strong coupling treatment done by Basso in~\cite{Basso:2011rs}  relies on this  rapidly convergent expansion~\footnote{It is worth noting that  the expansion for $\beta_1$ in $1/\sqrt\lambda$  can be identified with the GKP-loop expansion, i.e., the  coefficient  $b_{1,m}$ is a $m$-loop contribution.  The same no longer holds for $\beta_n$, $n>1$. In general, each coefficient $b_{n,m}$ mixes contributions from different loop orders.}. 
 It is now a simple exercise to check that, since $\beta_n$ starts at order $\lambda^{(2-n)/2}$, we have 
 \be
 \big(\Delta_P(j,\tau) - 2\big)^2 =  \tau^2 + 2\sqrt{\lambda} (j-2)\left( 1 +   \sum_{k=1}^{\infty}\,\lambda^{-\frac{k}{2}} H_k(j-2,\tau)\right),
   \label{eq:Basso.2}
 \ee
 with 
 \be
 H_k(j-2,\tau)=  \sum_{n=0}^{k}\,b_{n+1,k-n}(j-2)^{n}\,.
 \ee
 a polynomial of degree $k$ in    $(j-2)$. The form of this expansion should be compared with (\ref{ReggeonSpin}) for the function $j(\nu,\tau)$. Both expansions make explicit that we are doing a strong coupling expansion around the diffusion limit $j\to 2$ and $\lambda\to\infty$ with $\sqrt{\lambda}(j-2)$ fixed.

Let us now relate the expansions (\ref{eq:Basso.beta}) and (\ref{eq:Basso.2}) for the dimension $\Delta_P(j,\tau)$, to
the  expansion (\ref{ReggeonSpin}) for the Reggeon spin $j(\nu,\tau)$.  For simplicity we consider first the computation of the intercept.
Since $ \beta_1 (\lambda,\tau) >0$ the function $j(\nu,\tau)$ has a minimum at $\nu=0$ ($\Delta=2$). This minimum determines the Pomeron intercept $j_0=j(0,\tau)$. Thus, setting $\Delta_P(j_0,\tau) = 2$ in (\ref{eq:Basso.beta}) we learn that the 
intercept is fixed by 
\be
0=\big(\Delta_P(j_0,\tau) - 2\big)^2 = \tau^2+\beta_ 1(\lambda,\tau) (j_0 -2)   + \beta_2(\lambda,\tau)  (j_0-2)^2 + \beta_3(\lambda,\tau) (j_0-2)^3+ \cdots \,, 
\label{eq:S0}
\ee
with $\beta_n$ given by the expansion (\ref{eq:Basso.bn}).
Replacing the expansion for the intercept, as defined by  (\ref{ReggeonSpin}),
It is now possible to find  $j_0$  iteratively in an expansion
 \be
j_0 = 2 +  \frac{c_1}{\lambda^{1/2}} + \frac{c_2}{\lambda} + \frac{c_3}{\lambda^{3/2}} + \frac{c_4}{\lambda^2} + \cdots\,.
\label{eq:S0x}
\ee
Note that the coefficients $c_i$ are already defined in the expansion  (\ref{ReggeonSpin}) of the Reggeon spin function, 
more precisely we have $c_1=-\tau^2/2$ and $c_n=c_1 c_{n,0}$ for $n\ge 2$.
Substituting (\ref{eq:S0x})  and (\ref{eq:Basso.bn}) into (\ref{eq:S0}),  and collecting all terms in powers of $1/{\sqrt\lambda}$, one can determine $c_n$ iteratively.  To illustrate how this goes, we list here the first few coefficients (note that we set $b_{10}=1$),
\begin{align}
c_1&= - \tau^2/2  \,,\nn
c_2&= -b_{1,1}c_1 \,,     \nn
c_3&= -\big[ b_{1,1}c_2 + b_{1,2}c_1+ b_{2,0} c_1^2 \big] \,,\nn
c_4&= -\big[  b_{1,1}c_3  +b_{1,2}c_2+b_{1,3}c_1 + 2b_{2,0} c_1 c_2 +  b_{2,1} c_1^2 \big]\,,
 \label{eq:c-s}\\
c_5 &=
- \big[  b_{1,1}c_4  +b_{1,2}c_3+b_{1,3}c_2 + b_{1,4}c_1 + b_{2,0}(c_2^2+2 c_1 c_3) +  2b_{2,1} c_1 c_2  + b_{2,2} c_1^2+ b_{3,0} c_1^3\big] \,,\nn
c_6&=-\big[b_{3,1} c_1^3 + b_{2,3} c_1^2 + 3 b_{3,0} c_2 c_1^2 + b_{1,5} c_1 + 2 b_{2,2} c_2 c_1  + 2 b_{2,1} c_3 c_1 + 2 b_{2,0} c_4 c_1 + b_{2,1} c_2^2\nn
&\quad +\, b_{1,4} c_2 + b_{1,3} c_3 + 2 b_{2,0} c_2 c_3 + b_{1,2} c_4 + b_{1,1} c_5\big]\,.
\nonumber
\end{align}
 In the diffusion limit,  the intercept reduces to $\alpha_P(\tau)
 \equiv j_0(\tau)= 2-\tau^2/2\sqrt\lambda$, as stated above.  
 As for the leading twist   case, $\tau=2$, this intercept corresponds to the location of a square-root branch point for the spectral curve $\Delta_P(j,\tau)$, and it approaches $j=2$ in the limit of $\lambda\rightarrow \infty$.

More generally, we can  relate the coefficients of the polynomials  $H_k(j-2,\tau)$
entering the expansion (\ref{eq:Basso.2}) of $\Delta_P(j,\tau)$, with
 the coefficients of the polynomials $\tilde{j}_n(\nu^2,\tau)$
entering the expansion (\ref{ReggeonSpin}) of $j(\nu,\tau)$. These functions are simply related by the inversion formula
$\Delta_P\big(j(\nu,\tau),\tau\big) = 2+ i\nu$ given in (\ref{eq:BPST2}). This is a mechanical computation, so we only give here the 
relation between the first coefficients without further explanations (excluding the coefficients $c_{n,0}$ already given above)
\begin{align}
c_{3,1}&= \,b_{2,0}/2  \,, \quad\quad
c_{4,1}= \big[-3b_{1,1}b_{2,0}+b_{2,1}\big]/2\,,   \quad\quad c_{4,2}=0\,, 
\nn
c_{5,1}&=  \big[ 6 b_{1,1}^2 b_{2,0}-3 b_{1,2} b_{2,0}+2 \tau ^2 b_{2,0}^2-3 b_{1,1} b_{2,1}+b_{2,2}-\tau ^2 b_{3,0}\big]/2\,,  
\nn
c_{5, 2}&= \big[2 b_{2, 0}^2 - b_{3, 0}\big]/4\,,\quad\quad c_{5,3}=0\,, 
 \label{MatchCoeffs}
\\
c_{6, 1} &= \big[-10 b_{1, 1}^3 b_{2, 0} + 12 b_{1, 1} b_{1, 2} b_{2, 0} - 3 b_{1, 3} b_{2, 0} - 10 \tau^2 b_{1, 1} b_{2, 0}^2 +  6 b_{1, 1}^2 b_{2, 1} - 
3 b_{1, 2} b_{2, 1} +  4 \tau^2 b_{2, 0} b_{2, 1}\nn
&\quad - 3 b_{1, 1} b_{2, 2} + b_{2, 3} +  4\tau^2 b_{1, 1} b_{3, 0} - \tau^2 b_{3, 1}\big]/2 \,,\nn
c_{6, 2} &= \big[-10 b_{1, 1} b_{2, 0}^2 + 4 b_{2, 0} b_{2, 1} + 4 b_{1, 1} b_{3, 0} - b_{3, 1}\big]/4\,,
\quad\quad c_{6,3}=c_{6,4}=0\,.
\nonumber
\end{align}

\subsection{Explicit results for \texorpdfstring{${\cal N}=4$ SYM}{N=4 SYM}}

It is now clear that we can use the known results for the spectral curve $\Delta_P(j,\tau)$ for $j\sim 2$
to extract information about the strong coupling expansion Pomeron spin $j(\nu,\tau)$, and in particular to compute
the Pomeron intercept for $\tau\geq 2$.
We begin by noting that, since  $\beta_1(\lambda,\tau)= 2\tau\alpha_1(\lambda,\tau)$ is known analytically, it can be easily expanded in powers of   $1/\sqrt \lambda$, with the result~\cite{Basso:2011rs}
\be
\beta_1(\lambda,\tau) = 2 \lambda^{\frac{1}{2}} \Big(1  -\frac{1}{2\lambda^{1/2}} +\frac{4\tau^2-1}{8\lambda}+\frac{4\tau^2-1}{8\lambda^{3/2}}+\frac{-16\tau^4 + 104\tau^2 -25}{128\lambda^{2}} + \frac{-16\tau^4+56\tau^2-13}{32\lambda^{5/2}}+ \cdots\Big)\,.   
\label{eq:b1}
\ee
From this expansion it is straightforward to extract the coefficients $b_{1,m}$.
This is enough to fix $c_1$ and $c_2$ in the computation of the intercept (\ref{eq:c-s}), and in particular agrees with the diffusion limit
result  $c_1=-\tau^2/2$.~\footnote{The all coupling expansion carries more information about the coefficients $c_{n,k}$ of the $j(\nu,\tau)$
expansion (\ref{ReggeonSpin}). Indeed, it is simple to check that the combination 
$\sum_{k=0}^{n-2} (-1)^k \tau^{2 k} c_{n, k}$ is entirely fixed by the  coefficients $b_{1,m}$ with $m< n$.}

To find the intercept coefficients $c_n$ for $n>2$, knowledge of the 
coefficients $b_{n,m}$ for higher $\beta_n$ are required.     
As mentioned earlier, \eqref{eq:Basso.a2} is an expansion with increasingly divergent coefficients, i.e. $\alpha_n=O(\lambda^{n/2})$.  Clearly, very special  cancellations must take place in moving from (\ref{eq:Basso.a2}) to  (\ref{eq:Basso.beta}) for convergence.  Recently, explicit expressions for $\alpha_2(\lambda,\tau)$,  $\tau=2$ and $3$,  have been obtained~\cite{Gromov:2014bva}. With the aid of numerical analysis, together with  consistency matching with (\ref{eq:b1}), \cite{Gromov:2014bva} also gives strong coupling expressions for arbitrary 
$\tau$,  up to order $\lambda^{-3/2}$,
\begin{align}
\alpha_2(\lambda,\tau) &=  -\frac{\lambda}{2\tau^3} +\frac{\lambda^{1/2}} {2\tau^3}  + \frac{1}{4\tau}  + \frac{1-\tau^2 \big(24\zeta(3) +1\big)}{16 \tau^3 \lambda^{1/2}} - \frac{8\tau^4+ \tau^2 \big(72\zeta(3) +11\big) - 4}{32 \tau^3 \lambda}    \nn  
&\quad+\frac{24 \tau^4\big(16\zeta(3) + 20 \zeta(5) -7\big)  -48 \tau^2 \big(31\zeta(3) +20\zeta(5) +7\big)+ 75}{256 \tau^3 \lambda^{3/2}} + O(\lambda^{-2})\,.
   \label{eq:a2}
\end{align}
It is then possible to calculate expansions for $\beta_2$, with relevant low order coefficients $b_{2,m}$ extracted from the expansions of $\alpha_1$ and $\alpha_2$,
\begin{align}
\beta_2(\lambda,\tau)  
&=  2\,\left( \frac{3}{4}  - \frac{3 \zeta(3) -3/8}{2\lambda^{1/2}} - \frac{\tau^2 +9\zeta(3) - 5/8}{4\lambda^{3/2}}\right. \nn
&\quad+  \left.\frac{\tau^2\big(3\zeta(3) +15\zeta(5)/4  -27/16\big)-15\zeta(5)/2 - 93\zeta(3)/8 -3/16}{2\lambda^{2}} +  \cdots \right).   \label{eq:b2}
\end{align}
  The expansions of $\beta_n$ for $n>2$ are currently not known to high order. However, as discussed in section 6.3 of \cite{Gromov:2014bva}, from an analysis of classical energy, with semi-classical corrections, it is in principle possible to extract the $\tau$-independent coefficients $b_{n,0}$ and $b_{n,1}$ for all $n$. For our calculation of the intercepts we use $b_{3,0} =-3/16$ and $b_{3,1} = \big(60\zeta(3) + 60\zeta(5)-17\big)/32$.  Note that all coefficients  $b_{n,m}$ are polynomials in $\tau$, regular at $\tau=0$. We have also identified, for each coefficient, its order in a string loop-expansion~\cite{Gubser:2002tv,Frolov:2002av,Arutyunov:2003uj}.  With these coefficients it is possible to fix the $c_n$ up to $n=6$.
Thus, from (\ref{eq:c-s}), we find that the generalized Pomeron intercept is given by 
\begin{align}
\alpha_P(\lambda, \tau) = j_{0}(\lambda,\tau) = &\ 2 -   \frac{\tau^2 }{2\lambda^{1/2}} - \frac{\tau^2 }{4\lambda}  + \frac{\tau^2(-3+\tau^2) }{16 \lambda^{3/2}} - \frac{\tau^2 \left[-12 + \tau^2\big( 11+ 24\zeta(3)\big)\right]}{64\lambda^2} \nonumber \\ &+ \frac{\tau^2\left[ -63  + 6\tau^2\big(19+48\zeta(3)\big)-2\tau^4\right]}{256\lambda^{5/2}} \label{eq:intercept_tau}\\
&+ \frac{\tau^2\left[-216+\tau^2\big(637+1536\zeta(3)+480\zeta(5)\big)-2\tau^4\big(17+36\zeta(3)+60\zeta(5)\big)\right]}{512\lambda^3} + \cdots\,. \nonumber
\end{align}
For $\tau=2$ notice that the $\zeta(5)$ term is absent from $c_6$ due to cancellation, but it is in general present.
We will show in the next section that just by varying  $\tau$,  we can use this equation to calculate the Odderon intercept to the same order as above.

We may also compute, with the above information, the remaining coefficients $c_{n,k}$ up to $n=6$, by using (\ref{MatchCoeffs}). Such non-vanishing coefficients are~\footnote{Actually, since \cite{Gromov:2014bva} also computes $b_{4,0}=31/128$ and 
$b_{4,1}=\big(901-5520 \zeta(3) - 5120 \zeta(5) - 3640 \zeta(7)\big)/1024 $, we can determine
$c_{7,3}=391/1024$ and $c_{8,4}=\big(15081 - 27120 \zeta(3) - 12320 \zeta(5) - 3640 \zeta(7)\big)/8192$.}
\begin{align}
c_{3,1}&= \frac{3}{8}  \,, \quad\quad
c_{4,1}= 3 \,\frac{7 - 8 \zeta(3)}{32} \,,   \quad\quad c_{5,1}=  \frac{59 - 144 \zeta(3) - 2 \tau^2 }{64} \,,  \quad\quad c_{5, 2}= \frac{21}{64}\,,
\nn
c_{6, 1} &=  \frac{291 - 480 \zeta(5) - 76 \tau^2 - 
   48 \zeta(3) (32 + 7 \tau^2)}{256}   \,,\quad\quad 
c_{6, 2} =  \frac{137 - 204 \zeta(3)- 60 \zeta(5)}{128}\,.
\end{align}

\section{Conformal Odderon}\label{sec:Odderon}

It  is  appropriate  to begin by first mentioning  that the importance
of Pomeron lies partly in  the fact that all high energy hadron-hadron
total cross sections $\sigma_T$ continue to rise from collider to
cosmic ray energies. This universal behavior  can be understood as
driven initially  by the leading $1/N_c$  power law growth, $\sigma_T
\sim s^{\alpha_P -1}$  for the Pomeron intercept $\alpha_P >1$.
Eventual agreement with the Froissart bound $\sigma_T \sim log^2 s$
requires a re-summation of higher order terms in $1/N^2_c$ expansion~{\footnote{One often adopts an eikonal sum.  
Alternatively, the data is sometimes fitted directly by $\sigma_T \sim log^2
s$, the maximally allowed asymptotic term consistent with saturating
the Froissart unitarity bound.  A more thorough discussion can be found in \cite{Brower:2009} and references therein.}}. 
 The importance of the leading $C=-1$ component, generically referred to as the Odderon \cite{Lukaszuk:1973nt,Bialkowski:1974cp,Finkelstein:1989mf,Avila:2006wy}, lies in the fact that it contributes to the difference of the antiparticle-particle and particle-particle total cross sections,  $\Delta \sigma_T(s) \sim s^{\alpha_O-1}$.
 
In  the weak coupling limit,  the
Pomeron~\cite{Lipatov:1976zz,Kuraev:1977fs,BL,Lipatov:1985uk,KirschLipat}
can be associated  with 2-gluon exchange whereas Odderons  can be
thought as a $C=-1$ composite of a  three-gloun
system~\cite{Kwiecinski:1980wb,Wosiek:1996bf,Braun:1998fs,Bartels:1999yt,Ewerz:2005rg,Kovchegov:2003dm,Stasto:2009bc}.   Two leading
Odderons have been identified. One has an intercept slightly below
one~\cite{Kwiecinski:1980wb,Wosiek:1996bf,Braun:1998fs},
with $\alpha_{O,a}\approx 1-O(\lambda)$, and the second has an
intercept exactly at one,  $\alpha_{O,b}\approx  1$, up to third order in
the 't Hooft coupling~\cite{Bartels:1999yt}.  It has also been
suggested recently, for ${\cal N}=4$ SYM, that the latter remains exactly at $j=1$, to all orders in weak coupling~\cite{Bartels:2013yga,Kovchegov:2012rz}.  Interestingly, these correspond nicely with strong coupling analysis 
in the diffusion limit~\cite{Brower:2009,Avsar:2009hc}.  

Recall that, in a weak coupling BFKL  treatment, the spectral curve is obtained by an expansion about $j=1$ in the weak coupling $\alpha_s$, i.e., 
$j\approx 1-\alpha_s E_{N}\big(\Delta; \{\ell_N\}\big)$, where $E_N$ can be identified with the spectrum for a system of $N$-reggeon states~\cite{deVega:2002im,Derkachov:2002wz,Korchemsky:2003rc}, labelled by additional indices $\{\ell_N\}$. For $N=2$, this leads to 
\be
j=1 +\alpha_s\Big \{ 2 \Psi(1) - \Psi\Big(\frac{(n+3)-\Delta}{2}\Big) - \Psi\Big(\frac{\Delta-(n+1)}{2}\Big) \Big\} 
\ee
where there is  a single index,  $n=0,1,\cdots$, labelling the principal series  representation of $SL(2,{\cal C})$.  The leading Pomeron corresponds to $n=0$.  Note that this representation is a perturbation about $j=1$,  and  the right-hand side develops singularities   due to poles of the $\Psi$-function.  This representation therefore cannot be extended to  the region of large $j$ and $\Delta$. For $N=3$, appropriate for the Odderons, one solution coincides with that for $N=2$, with $n=1$ and $\Delta=2$, leading to an Odderon  intercept $j=1$, as indicated above.

In this Section, we examine these strong coupling results, going beyond the diffusion limit. In particular,   we clarify how for the special Odderon solution,  $\alpha_{O,b}= 1$ at $k=0$, can hold to all order in $1/\sqrt\lambda$. 

For the Odderon, the large $\lambda$ difusion limit corresponds to setting $\lambda\to\infty$ and $j\to 1$, with $\sqrt{\lambda} (j-1)$ fixed.
In this limit the Odderon propagator can be obtained  by perturbing about the EOM for the anti-symmetric Kalb-Ramond field, 
$B_{MN}$ in $AdS$:
$(-\square_{Maxwell} + m_{ads}^2)B_{MN}=0$.  
Here  $\square_{Maxwell}$ stands for the Maxwell operator.  Its exact
form can be found in \cite{Brower:2009}, and it can again be
diagonalized in terms of $O(4,2)$ Casimir.  The result is that  
all modes with odd positive integral $j$ contribute to the exchange,  and one arrives at an effective propagator in the 
$\nu^2-j$ plane of the form
\begin{equation}
G_O(j,\nu) \sim   \frac{1}{ \nu^2+ m^2_{AdS}      +   2\sqrt\lambda (j-1)  } \,. 
 \label{eq:Odderonpropagator}
\end{equation}
As for the Pomeron, the relevant string modes can be represented by
on-shell world-sheet Reggeon vertex operators ${\cal V}_O^\pm$ in $AdS$.
The on-shell condition, $L_0{\cal V}_O^\pm=\bar L_0{\cal V}_O^\pm={\cal V}_O^\pm$, in analogy with (\ref{eq:BPST2}),  leads to 
$\Delta_O(j) = 2 + i \nu \label{eq:BDT.O}$.
It follows  that  the Odderon spectral curve in the strong coupling diffusion limit,  given   by the pole locations of  $G_O(j,\nu)$, is  
\be
\big(\Delta_O(j)-2\big)^2= m^2_{AdS}       +   2\sqrt\lambda (j-1)\,. 
\label{eq:D.O1}
\ee
  Denoting   $\alpha_O(\lambda)$ for  the Odderon intercept, it follows that
\be
\Delta_O(j)=2 + \sqrt 2 \lambda^{1/4} \sqrt {j-\alpha_O} \,, 
\label{eq:singularity.O}
\ee
where in the diffusion limit
\be
\alpha_O(\lambda) = 1- \frac{m^2_{AdS} }{2\sqrt\lambda}\,. 
\label{eq:O-Intercept}
\ee
We  also stress that because of super-symmetry  the anomalous  dimension at $j=1$ is zero, more precisely,
$
\Delta_O(1)=2+m_{AdS}
$, 
for any value of the coupling. In the diffusion limit, the spectral curve $\Delta_O(j)$ is again parabolic.

From the SUGRA modes, we see that there are two sets of solutions, a set with $m^2_{AdS,a}=(4+k)^2$ which we dub as set (a), and 
a set (b) for which  $ m^2_{AdS,b} = k^2$  with   $k=0,1,2,\cdots$.   For $k\neq 0$, these modes can be associated with fluctuations in $S^5$. For the $k=0$ mode of set (b), at $j=1$,  it is known that it  can be gauged away since its coupling is through the field-strength which vanishes~\cite{Brower:2000rp,Brower:1999nj,Kim:1985ez}. However, in the diffusion limit of $\lambda$ large but finite,  one moves away from  $j=1$ with  the perturbation introducing 
 an effective $AdS$ mass so that the field-strength no longer  vanishes and the mode is now  physical. 
We will return  later  to discuss this mode further.  For each mode $k$ of the $S^5$,
the two distinct  $AdS$ masses directly lead to two distinct Odderon trajectories with intercept given by
(\ref{eq:O-Intercept}) with the associated mass, 
\be
\alpha_O(\lambda) = 1- \frac{m^2_{AdS,a/b} }{2\sqrt\lambda}\,. 
\label{eq:odderon_diffusion.ab}
\ee

From the perspective of OPE, at $j=1$, there are two sets of  conformal primaries, each indexed by integer $k$, with \emph{protected conformal dimensions}~\footnote{Of course, each contributes only to correlation functions with appropriate R-charge.},
\begin{align}
\Delta_O^{(a)}(1) &= 2+ \tau_{a} = 2 + m_{AdS,a}=6+k\, ,  \label{eq:tau.a}\\
\Delta_O^{(b)}(1)& = 2+\tau_b= 2  + m_{AdS,b}=2+k  . \label{eq:tau.b}
\end{align}
Candidate dual CFT operators dual to these protected  string modes are $Tr(F_{\perp\pm}F^2 Z^k) + \cdots$ and $Tr(F_{\perp\pm} Z^k)$, respectively. As usual, one expects that higher spin operators in the leading Regge trajectory can be  obtained by acting with derivatives $D^\pm$, leading to operator dimensions $\Delta_O^{(a)}(2n+1)$ and $ \Delta_O^{(b)}(2n+1)$ respectively, at $j=2n+1$, $n=1,2\cdots$.  However, these can only be obtained meaningfully beyond the diffusion limit.

\subsection{Odderon Intercepts in Strong Coupling}
\label{sec:Basso-O}

To go beyond the diffusion limit, let us return to 
(\ref{eq:Basso.beta}).
In analogy to that equation and (\ref{eq:Basso.2}), we expand $\big(\Delta_O(j,\tau) - 2\big)^2$ about $j=1$, 
\be
\big(\Delta_O(j,\tau) - 2\big)^2= \tau_O^2 + \beta^{(-)}_1 \, (j-1)+ \beta^{(-)}_2 \, (j-1)^2+ \beta^{(-)}_3 \, (j-1)^3 + \cdots \, .
\label{eq:Basso.4.b}
\ee
We have also added a superscript to the expansion coefficients, $
\beta^{(-)}_n$,  to remind ourselves that we are dealing with the  $C=-1$ sector.
Recall that (\ref{eq:Basso.4.b}) properly reflects the symmetry in $\Delta_O \leftrightarrow 4-\Delta_O$.  To match the diffusion limit, we  require
\be
\beta_1^{(-)}(\lambda,\tau_O)=2\sqrt\lambda + O(1)   \, .
\ee
As is the case with the Pomeron,    $j(\Delta_O)$ has a minimum at
$\Delta_O(\alpha_O,\tau) =2$, which defines the   Odderon intercept. This, of course, is also equivalent to the existence a square-root singularity, (\ref{eq:singularity.O}).
We further assume that, as the case for the Pomeron, in the strong coupling limit, the radius of convergence for (\ref{eq:Basso.4.b}) is $O(\lambda^{1/2})$, and, $\beta^{(-)}_n=O(\lambda^{(2-n)/2})$.  Correspondingly, we can develop a systematic expansion for  $\beta_n^{(-)}(\lambda)$ in $1/\sqrt\lambda$, 
\be
\beta^{(-)}_n=2\lambda^{\frac{2-n}{2}}\left( b^{(-)}_{n,0} + \frac{b^{(-)}_{n,1}}{\lambda^{1/2}}+ \frac{b^{(-)}_{n,2}}{\lambda}+ \frac{b^{(-)}_{n,3}}{\lambda^{3/2}}+\cdots\right).  
\label{eq:Basso.bn.O}
\ee

We are now in the position to  carry out a similar analysis for  Odderon intercepts beyond the diffusion limit. Consider the expansion for the intercept
\be
\alpha_0(\lambda)= 1 + \frac{c_1^{(-)} }{\lambda^{1/2}}+ \frac{c^{(-)}_{2} }{\lambda} +\frac{c^{(-)}_{3} }{ \lambda^{3/2}}+\frac{c^{(-)}_{4} }{\lambda^2}+\frac{c^{(-)}_{5} }{\lambda^{5/2}} +\cdots.
\label{eq:S-c}
\ee
The coefficients in this expansion  can be found by solving $\Delta_O(\alpha_O,\tau) = 2$ iteratively.
Observe that  the situation in nearly identical to that for the Pomeron. It follows that $c_i^{(-)}$ are given exactly by the corresponding coefficients $c_i$ for the Pomeron intercept,  Eq. (\ref{eq:c-s}), with the replacements of $\tau$ by $m_{AdS}$, $c_i$ by $ c_i^{(-)}$ and  $b_{n,i}$ by $ b^{(-)}_{n,i}$.
For simplicity, we shall drop the superscript  in what follows.

 In the diffusion limit, our two sets of Odderon solutions are
 structurally similar. However, there is no particular reason why
 these two sets remain similar in higher orders and we shall treat
 them separately in what follows.  
We shall  first consider type-(a), characterized by $\tau_a=m_{AdS}=4+k$, $k=0,1,2,\cdots,$  before treating the case for type-(b). It is worth mentioning again that the all-coupling formula (\ref{eq:Basso.a1})  was derived from ABA equations, without the so-called wrapping corrections. It is surprising that they do not appear in the present context as one would expect them especially in the small spin limit.  The physical motivation for their absence is not fully understood~\footnote{See \cite{Basso:2012ex,Gromov:2012eg} for a discussion and comparison to ABJM theory.}. Furthermore, it is supposed  to hold only for the $sl(2)$ sector for the configuration with ``minimum mode numbers", which should correspond to the spectral curve with minimum scaling dimension. It is indeed possible to generalize the solution of  ABA for ``non-minimum string modes". However, no systematic attempt has been made \cite{Basso:2012ex,Gromov:2012eg}. Further discussion will be provided in Sec. \ref{sec:Discussion}.

 \noindent {\bf I -- Type-(a) Odderons:}

 As pointed out earlier, the $j=1$ mode survives in the supergravity limit, and it can be identified with the protected CFT operator   of the type   $Tr(F_{\perp\pm}F^2 Z^k) + \cdots$. This conformal primary, just as the case for the Pomeron, can formally be considered as a descendent of  super-conformal primary in  the $sl(2)$ sector.  However, there can be many spectral curves emanating from this protected configuration at $j=1$, for $\lambda $ finite. We assume that   the all-coupling formula (\ref{eq:Basso.a1}) for the slope function applies to the set $\Delta_{O,a}$, corresponding to the ``minimum twist" set, and  
we shall proceed  to calculate the Odderon intercept to higher orders in $1/\sqrt\lambda$ under this assumption.  The validity of this assumption  will be examined  in the next section. 

\begin{figure}
		\begin{center}
		   	\includegraphics[width=0.6\columnwidth,keepaspectratio]{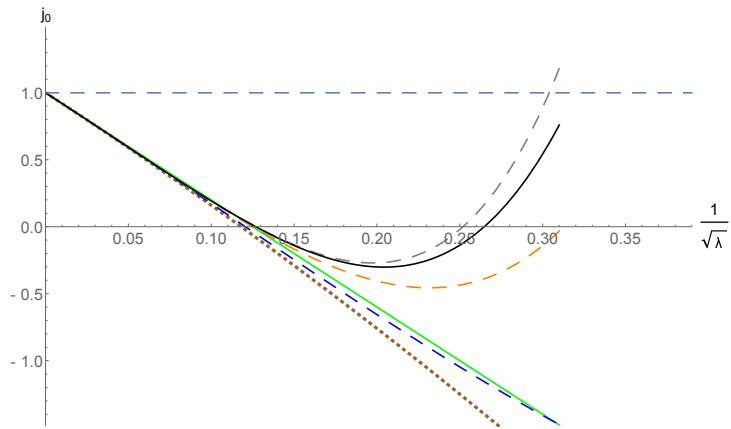}
   	\caption{\label{fig:ointer}\footnotesize{Odderon-a intercept
            at strong coupling.  The solid  green-line is to first
            order in $1/\sqrt \lambda$, the dotted brown-line is to
            second order, the dashed blue-, orange- and grey-line are
            to third order, fourth order and  fifth order
            respectively. Finally the solid black-line is the
            intercept up to sixth order.}}
   	\end{center}
\end{figure}

\begin{figure}
		\begin{center}
   	\includegraphics[width=0.6\columnwidth,keepaspectratio]{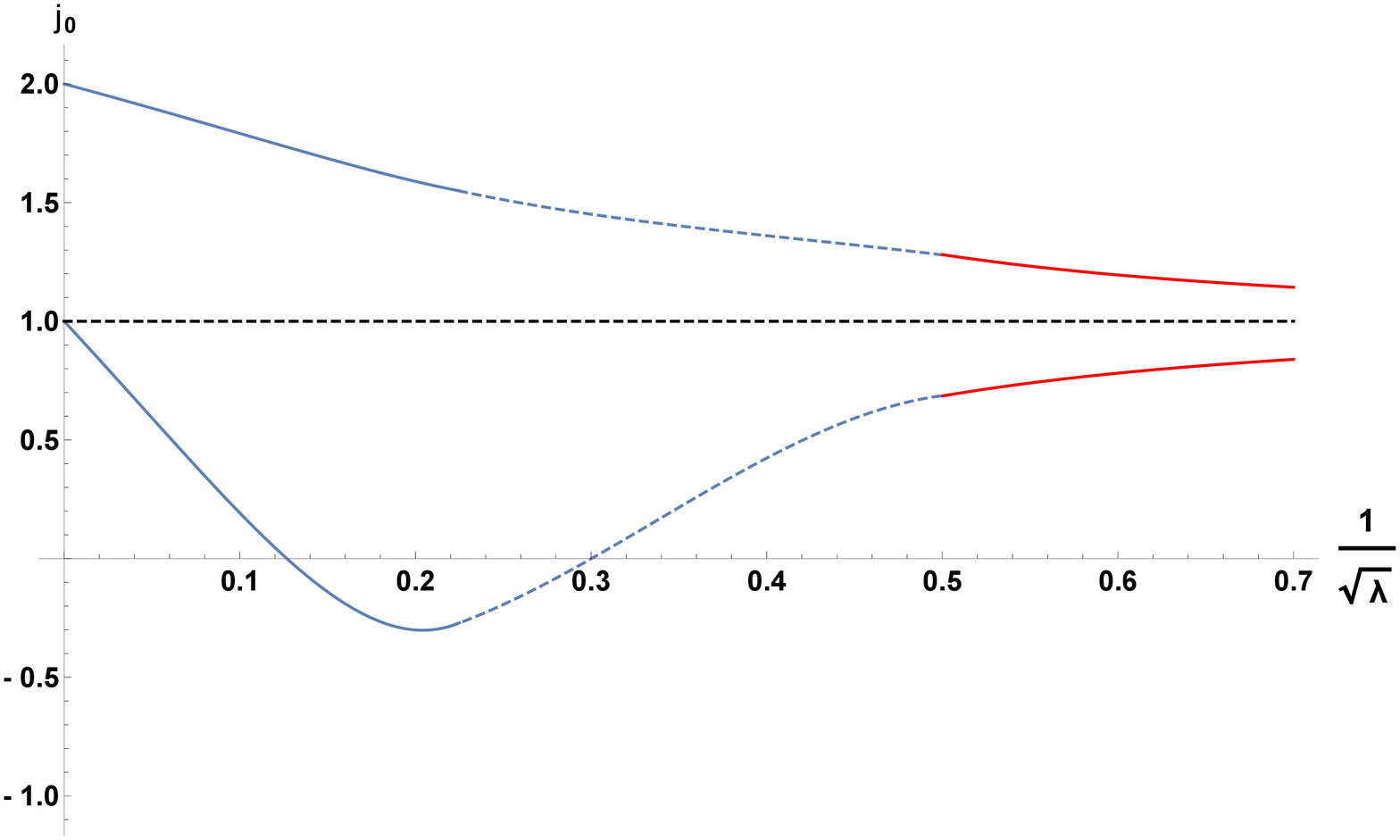}
   	\caption{\label{fig:ointer-2}\footnotesize{The Pomeron and Odderon-(a) intercepts from strong to weak coupling.  The dark blue curves are the calculated strong coupling results, the red curves are the known weak coupling intercepts, and the dashed line is an interpolation.  It is interesting to note that up to their current orders, both the Pomeron and Odderon intercepts appear consistent with weak coupling results in the transition region. Black  dashed  line is for the Odderon-(b) solution where $\alpha_{O,b}=1$.}}
   	\end{center}
\end{figure}

That is, we assume that, for type-(a) Odderons,  $\beta_n(\tau_{(a)})=\beta_n(4+k)$,  with  $\beta_n$ given by the same functions used for the case of Pomeron.  With this we can find the intercept for any $k$, but for simplicity, and to study the case most relevant for QCD, we write the result in the limit $k\rightarrow 0$. Hence $\tau_{(a)} =4$, and we take advantage of the expansion \eqref{eq:S-c}.
 Under the  assumptions made above, $c_i$ can be found  by solving the equation
\begin{equation*}
0= 4^2 + \beta_1(\lambda,4)  (j_0-1) + \beta_2(\lambda,4)  (j_0-1)^2  + \beta_3(\lambda,4)  (j_0-1)^3  + \beta_4(\lambda,4)  (j_0-1)^4   + \cdots \,,
\end{equation*}
iteratively.   It is possible to directly adopt the calculation previously done  for the Pomeron intercept in Sec. 3, with $c_n$ given by (\ref{eq:c-s}), by evaluating equation \eqref{eq:intercept_tau} at $\tau=4$, after shifting the spin $j$ by 1. 
One finds
\be
\alpha_{O,a}=1 - \frac{8 }{\lambda^{1/2}}- \frac{4 }{\lambda} +\frac{13 }{ \lambda^{3/2}}+\frac{96\zeta(3)+41}{\lambda^2}+ \frac{288\zeta(3)+\frac{1249}{16}}{\lambda^{5/2}} +\frac{-720\zeta(5)+192\zeta(3)+\frac{159}{4}}{\lambda^{3}}  +\cdots. \label{eq:dualconformalOdderon.a}
\ee
This  intercept $\alpha_{O,a}$  is illustrated in Fig. \ref{fig:ointer}. Note that coefficients $\{c_n\}$ change signs,  with $c_1,c_2<0$,  $c_3,c_4,c_5>0$, and $ c_6<0$.  We also  note that, in the range $0<1/\sqrt\lambda <0.3$, where strong coupling is expected to be useful, the intercept $\alpha_{O,a}$ is below $j=1$. As one increases $1/\sqrt\lambda$ beyond $0.2$,  interestingly, it  begins to turn around and move towards $j=1$, as it should, only after $c_4$ and higher terms are kept. Note as well that the intercept does not continue to blow up, but begins to flatten out as higher orders are taken into account, e.g., with $c_6<0$.  This behavior fits nicely with the expected matching behavior to first order weak coupling calculation, at $1/\sqrt \lambda \approx 0.3$, as shown in Fig. \ref{fig:ointer-2}

Notice that it is a simple exercise to determine the coefficients in an expansion of the type (\ref{ReggeonSpin}) for the 
Odderon spin function $j(\nu,\tau)$ .

\noindent {\bf II -- Type-(b) Odderons  and  $\tau_b=k\rightarrow 0$:}

As stated earlier, although there are no obvious  structural
differences between type-(a) and type-(b) Odderon spectral curves in
the diffusion limit, one crucial difference is the absence of  a
protected local CFT operator at $j=1$ at $\tau_b\rightarrow  0$,  in
the supergravity limit.  For $k\neq 0$, the simplest set of local operators one can identify are $Tr[(D^\pm)^{j-1} F_{\pm\perp}Z^k]$.  These operators decouple however in the $k\rightarrow 0$ limit.  It is likely that the lowest physical mode on
the $k=0$ spectral curve occurs at $j=3$ is
$Tr[F_{\pm\perp}F_{\pm\perp}F_{\pm\perp}]$ and it is not protected. It has been suggested~\footnote{B. Basso, private communication.} that, for $j=3,5,\cdots$,  these type-(b) modes could be associated with local operators: $Tr[ D_\pm^{j-3} F_{\pm\perp}F_{\pm\perp}F_{\pm\perp}]+\cdots$.   However, these lead to a different system of $sl(2)$ spin chain, and a separate  analysis is  required~\cite{Belitsky:2004cz,Belitsky:1999bf,Beccaria:2007pb,Beccaria:2011kv}. An equally interesting possibility is to consider the sequence $Tr[ (F_{\pm\perp}F_{\pm\perp})^{(j-1)/2}F_{\pm\perp}]$.  Neither sequence leads to well-defined local operators at $j=1$.  For this and other considerations, we do not expect the  all-coupling formula (\ref{eq:Basso.a1}) to work for type-(b) Odderon, specially for the $k=0$ mode. Therefore, we shall proceed to carry a more general analysis {\it without}  assuming the Basso formula for the slope function, $\alpha_1$. We shall search for, if any, universal behavior which might survive in the limit of $\tau_b\rightarrow 0$.

We proceed by making the expansion for $\beta_n$,  (\ref{eq:Basso.bn.O}),  as before, and arrive at 
\begin{align}
\big(\Delta_{O,b}(j,\tau_b) - 2\big)^2 =  \,\, \tau_b^2      
&+  2\,\lambda^{1/2} \left(1-\frac{b_{1,1}}{\lambda^{1/2}}+\frac{b_{1,2}}{\lambda^{3/2}}+\frac{b_{1,3}}{\lambda^{2}}+\cdots\right)\,(j-1) \nn
& +  2\, \left(b_{2,0}+\frac{b_{2,1}}{\sqrt{\lambda}}+\frac{b_{2,2}}{\lambda}+\frac{b_{2,3}}{\lambda^{3/2}}+\cdots \right)\,(j-1)^2\nn
& + 2\,\lambda^{-1/2}\left(b_{3,0}+\frac{b_{3,1}}{\lambda^{1/2}}+\frac{b_{3,2} }{\lambda}+\frac{b_{3,3}}{\lambda^{3/2}}+\cdots \right)\,(j-1)^3   \nn
&+ O\big((j-1)^4\big)\, . \label{eq:D-b}
\end{align}
Note, with the exception of $\tau_b=k$ and $b_{1,0} = 1$, other coefficients $b_{n,j}$ are {\it unspecified}.  
Expanding $\alpha_O$ as in (\ref{eq:S-c}), after substituting into (\ref{eq:D-b}), the coefficients $c_i$ can be determined iteratively, e.g., leading to  formulas essentially given by  Eq. (\ref{eq:c-s}).  To be more explicit, we can directly express all coefficients $c_i$ in terms of $b_{n,i}$. 
For the first few terms we
have, from  (\ref{eq:c-s}),
\begin{align}
c_1(\tau_b)  &=  -\frac{\tau_b^2}{2}  \, ,\nn
c_2(\tau_b) &=  \frac{\tau_b^2}{2}\,b_{1,1} \, ,\nn
c_3(\tau_b)& =  \frac{\tau_b^2}{2}\left(     \left(-b_{1,1}^2 + b_{1,2}\right) - b_{2,0}\,\frac{\tau_b^2}{2} \right), \nn
c_4(\tau_b)&= \frac{\tau_b^2}{2}\left(   \left(b_{1,1}^3 -2b_{1,1}b_{1,2}+b_{1,3}\right) +\left(3b_{1,1}b_{2,1}-b_{2,1}\right)\frac{\tau_b^2}{2} \right),\\
c_5(\tau)&= \frac{\tau_b^2}{2}\left(   \left(- b_{1,1}^4+\cdots\right) +\left(-6b^2_{1,1}b_{2,0}+\cdots \right)\frac{\tau_b^2}{2}+\left(-2b^2_{2,0}+\cdots\right)\frac{\tau_b^4}{2^2}\right), \nn
c_6(\tau_b) &= \frac{\tau_b^2}{2}\left(  \left(b_{1,1}^5+ \cdots\right) +\left(10b^3_{1,1}b_{2,0}+\cdots \right)\frac{\tau_b^2}{2}+\left(10b_{1,1}b^2_{2,0}+\cdots\right)\frac{\tau_b^4}{2^2} \right). \nonumber
\end{align}
From~\cite{Gromov:2014bva},  we  expect that $b_{n,j}$ to be  polynomials of $\tau_b^2$. Note that in the limit $\tau_b\rightarrow 0$, all coefficients vanish as $\tau_b^2$, e.g., $c_1\sim \tau_b^2$, $c_2\sim \tau_b^2$, etc., for {arbitrary values for coefficients $b_{n,i}$.    It is also easy to verify that,  $c_{n+1}/c_n \sim O(1)$, for $\tau_b\rightarrow 0$.  We thus arrive at an important  result where, $c_n =O(\tau_b^2)\rightarrow 0$, for all $n$, in the limit $\tau_b\rightarrow 0$. It follows that the leading Odderon intercept for the set-(b), for $\tau_b=0$,  remains at 
\be
\alpha_{O,b}=1\, , \label{eq:Ode-b0}
\ee
without higher order correction in an $1/\sqrt\lambda$ expansion.  This is the long  promised  result.  To state it more graphically, in the limit $\tau_b\rightarrow 0$, higher order corrections can change the shape of the spectral curve, without changing its minimum at $\Delta=2$.

For $\tau_b\neq 0$, more information is required in order to determine
the higher order expansion for their intercepts, e.g., adopting the
all-coupling formula (\ref{eq:Basso.a1}). We will not engage in this exercise here, but
  note that due to the generality of the above derivation, the $\tau_b = 0$ result would survive for all possible expansions of $\beta_n$.

\section{Conclusions}\label{sec:Discussion}

We have   focussed in this study  on the  leading $C=\pm 1$ Regge
singularities, the  Pomeron and Odderon respectively, in strong coupling. 
Central to our discussion is the notion of spectral curve $\Delta(j)$ for single-trace gauge invariance operators of ${\cal N}=4$ SYM.   Identifying $\Delta(j)$  in weak coupling  remains  involved due to possible operator mixings~\cite{Belitsky:2004cz}. 
In strong coupling,   spectral curves for leading twist can be identified    with bulk degrees of freedom for  $D=10$ SUGRA on $AdS_5\times S^5$~\cite{Kim:1985ez}.   The Pomeron trajectory can be associated with the Reggeized Graviton,   while the Odderon trajectories   correspond to  Reggeized anti-symmetric $AdS_5$ Kalb-Ramond tensor-fields.   With   string-vertex operators~\cite{ Brower:2006ea,Brower:2009},    
the relevant single-trace CFT operators  and their associated  string modes  for both Pomeron and Odderon sectors can be identified. 

We began by first  providing  a general discussion on  Regge limit  in CFT  from the perspective of light-cone OPE, and  showed  how the double-Mellin transforms (\ref{eq:newgroupexpansion}) and (\ref{MellinJ})  can be used  directly in a Minkowski setting.
A Regge dictionary~(\ref{eq:CFT-Regge-dictionary}) is established
between CFT in coordinate representation and that based on AdS/CFT in
a momentum treatment.  An important and probably difficult theoretical
problem left unresolved is to determine the conditions required in a
conformal theory to allow for this representation in the double-Mellin
plane.  Is this conformal Regge representation a generic property of
all 4D Lorentz invariant conformal theories, or is it restricted  to  a smaller
class of theories?

Due to integrability, these  spectral curves   can in principle be determined~\cite{Beisert:2010jr}, with their inverse  $j(\Delta)$ being  symmetric under $\Delta  \leftrightarrow  4-\Delta$,  due to   conformal invariance.   
 In this study, we have focussed on ``short strings" where  each spectral curve  takes on a relatively simple form in the large $\lambda$ limit while maintaining the  $\Delta  \leftrightarrow  4-\Delta$ symmetry.   In particular, by adopting the  approach advocated in~\cite{Gromov:2011bz,Basso:2011rs,Basso:2012ex,Gromov:2012eg}, higher order expansion in $1/\sqrt\lambda$ for the Pomerom intercept has been carried out recently~\cite{Costa:2012cb, Kotikov:2013xu,Gromov:2014bva}. We have generalized this  analysis for the Pomeron sector to include non-zero R charge, (\ref{eq:intercept_tau}),  and have also extended the treatment to the case of Odderons, (\ref{eq:dualconformalOdderon.a}) and (\ref{eq:Ode-b0}).
For the case of the Pomeron with large classical R charge, it would be interesting to see the appearance of this trajectory in the Regge limit of
four-point functions of heavy operators computed in  \cite{Caetano:2012ac} at strong coupling. 

It is important to emphasise that our analysis has been carried out in the context of AdS/CFT, appropriate for a strong coupling  expansion in the large-$N_c$ limit.  Simplicity in the complex $\Delta-j$ plane is achieved  partly  due to the ability to identify modes of SUGRA with protected gauge-invariant YM operators   in the limit of $\lambda\rightarrow \infty$, as discussed in Sec. \ref{sec:intro} and also in Sec. \ref{sec:Basso-O}.   This in turn allows us to treat leading $\Delta(j,\tau)$ curves which  dominate  the Regge limit through the double-Mellin representation discussed in Sec. \ref{sec:OPE}. It is expected~\cite{Brower:2006ea} that additional sub-dominate spectral curves exist, leading to ``fine-structure" to the complex $\Delta-j$ plane.  It is interesting to note in this connection that anomalous dimensions of higher-twist Wilson operators in generic gauge theories have previously been investigated and a robust structure, particularly at large $j$, has been found, e.g., a band of trajectories of width growing logarithmically with  spin-$j$ \cite{Belitsky:2008mg}.   At large-$j$,  anomalous dimensions increase with spin as $\ln j$, with leading coefficients given by ``cusp-anomalous dimensions". The analysis in \cite{Belitsky:2008mg} was carried out for physical integral $j$-values, in the framework of asymptotic Baxter equation and also based on semiclassical expansion.  It is reasonable to expect  that this ``band-like" structure identified for higher-twist sectors should persist at low-$j$ values, and it is interesting to ask how a smooth connection can be achieved~\footnote{It is also appropriate to point out  that,  in the case of a theory with a mass gap, the analytic continuation in $j$ is unique for the partial-wave amplitudes, following what is  known as  the ``Froissart-Gribov" procedure. We fully expect a similar procedure can be carried out for generic CFTs, e.g.,  by  adopting, for instance, the Mack representation~\cite{Mack:2009mi}  as a starting point of discussion. As a consequence, there is a unique analytic continuation for $\Delta(j,\tau)$ away from integral $j$ and $\tau$ values. This will be discussed  in a  future treatment.}.  Clearly, this can only be discussed  meaningfully in the context of the large-$N_c$ limit where the $j$-plane structure is expected to be simplified, e.g., adopting  the approach  of quantum spectral curves,  advocated in~\cite{Alfimov:2014bwa,Gromov:2014bva}.   More immediately, the all coupling analysis in \cite{Gromov:2011bz,Basso:2011rs,Basso:2012ex,Gromov:2012eg}, which focuses on the small spin region, and is also based on the asymptotic Bethe ansatz, can shed light on this issue.

Since our result depends crucially on the small spin expansion (\ref{eq:Basso_expansion}), 
it is worth first adding a brief comment  on the  slope function $\alpha_1(\lambda,\tau)$,  (\ref{eq:Basso.a1}), as promised earlier.   We first note that  the set of  gauge invariant operators in the $sl(2)$ sector, designated symbolically by  $ Tr (D_+^SZ^\tau) $, should  be interpreted as a collections of operators, 
$
Tr (D_+^{s_1}Z D_+^{s_2}Z\cdots D^{s_\tau}_+Z)
$,
with $\sum_{j=1}^\tau s_j=S$.   In the large $N$ limit, the dilation operator closes on this subspace,  leading to a set of spectral curves, $\Delta_{Z,k}(S,\tau) $,  labelled   by an index $k$.
Our focus here is for strong coupling, where, in the diffusion limit,  $\Delta_{Z}\approx \tau +   (\sqrt \lambda/\tau)S$. Therefore, {\it all these curves are degenerate in this limit.  }  
 The degeneracy is lifted by going beyond the leading $1/\sqrt\lambda$ limit.
Expanding  $\Delta_{Z,k}(S,\tau) $ for $S$ small, i.e., $\Delta_{Z,k}(S,\tau)= \tau + \alpha_{1,k}(\tau,\lambda) S +O(S^2)$, these slope functions,  $\alpha_{1,k}(\tau,\lambda) $,  can be found via ABA, without ``wrapping corrections", where   the index $k$  can be  specified by a set of  filling fractions~\cite{Basso:2011rs,Basso:2012ex,Gromov:2012eg}~\footnote{To be more explicit,  $k\equiv \{\kappa_m\}$, where $\sum_{m\neq 0} \kappa_m=1$.
Since  $\Delta_{Z}\approx \tau +  (\sqrt \lambda/\tau) S$  in the diffusion limit, it follows that there exists another constraint
$
\sum_m |m|\kappa_m=1.
$   Minimal filling corresponds to $\kappa_1=\kappa_{-1}=1/2$.}.
The ``minimum filling", (or ``minimum mode" configuration),  leads to the original  Basso formula.  
Clearly, there are many additional  lower spectral curves, corresponding to other allowed filling fractions. It would be interesting if  these additional spectral curves  can  be identified with those found  at high-$j$ and at weak coupling~\cite{Belitsky:2008mg,Belitsky:2004cz,Belitsky:1999bf,Beccaria:2011kv,Beccaria:2007pb}.
It is also interesting to find out  how these can be related to the mode number in the discussion of GKP strings.
 In our current treatment, the minimum filling solution has been used for both the Pomeron and the type-(a) Odderon sectors. Note that for both the Pomeron and the type-(a) Odderon keeping the $O(1/\sqrt{\lambda})$ result for the intercept stemming from the Basso formula agrees with the Pomeron and Odderon intercepts found independently by perturbing about the supergravity limit \cite{Brower:2006ea,Brower:2009}.  
}

Let us turn next to   the type-(b) Odderon.  As we have stated earlier, in the super-gravity limit, the $k=0$ mode decouples  at $j=1$, and, for $\lambda$ finite, its physical modes begin at $j=3, 5,\cdots$. In Sec. \ref{sec:Basso-O}, we have carried out a more generally analysis without invoking the 
all-coupling formula (\ref{eq:Basso.a1}). Our treatment is based on the  structure of large $\lambda$ expansion. Given  the diffusion result of $\alpha_{O,b}=1$, for $k=0$,  Eq. (\ref{eq:Ode-b0})    follows to all orders.   The phenomenon  of  decoupling also  has its counter part  in flat space string theory.   For $d=4$,  Kalb-Ramond tensor field does not lead to a spin-1 massless particle since $B_{\mu\nu}$ has only one independent transverse  component.  As a 4-d Regge trajectory, its higher string recurrences at $j=3,5,\cdots$ are  physical.  Therefore, the issue of decoupling  can be accomplished by an appropriate  vanishing of the coupling,  while the dynamics of the whole trajectory remains.  (Instead, at $t=0$, $B_{\mu\nu}$ leads to a spin-0 state, the ``axion".)   A more detailed analysis will be presented elsewhere.

In this study, we have not fully explored  the consequence of super-symmetry, in particular the possibility of more general symmetry patterns for   spectral curves  $\Delta_Z(S,\tau)$. A useful study is a  careful examination for the spectral curves for other modes of SUGRA, e.g.,  scalars, vectors, etc. A preliminary finding involves the possibility of having a more complex structure, e.g., the symmetry about $\Delta=2$ is realized by a pairing of spectral curves, with one symmetric about $\Delta=0$ and the other about $\Delta=4$.  Equally interesting is the question for the Odderon intercept in strong coupling  from a brane construct alone, without imposing super-symmetry, and its relation to results obtained in weak coupling.

 As stressed in \cite{Basso:2012ex}, $\Delta_Z(S,\tau)$  is in general a complicated function of $S$, $\tau$, $\lambda$, and also of other quantum numbers. It can in principle possess infinitely many branches,  connected through the so-called ``crossing-point singularities,"  i.e., the phenomenon of level-crossings, leading to root-type branch point in $\Delta_Z(S,\tau)$.  Indeed, in perturbing about the supergravity limit, the multi-valued property of $\Delta_Z$ seems to play an important in ensuring $\Delta \rightarrow \Delta-4$ symmetry. It should be stressed that these crossing-point singularities do not lead to branch points for the analytically continued conformal partial-wave amplitude, Eq. (\ref{eq:groupexpansion}).
This type of crossing-point singularities has also been noted previously in a weak coupling treatment for Odderons via BFKL-like analysis.   As mentioned earlier in Sec. \ref{sec:Odderon}, in  such a treatment, the spectral curve is obtained by an expansion about $j=1$, i.e., $j\approx 1-\alpha_s E_{N}(\Delta; \{\ell\})$, where $E_N$ can be identified with the spectrum for a system of $N$-reggeon states~\cite{deVega:2002im,Derkachov:2002wz,Korchemsky:2003rc}. It can be shown that, for $N\geq 3$, level-crossing occurs, leading to crossing-point branch points. However, it is unclear if  there is a  correspondence for such singularities at  weak and strong coupling. Making a precise connection between the  strong and weak coupling Odderon solutions remains a challenge~\footnote{There also exists a slight disagreement~\cite{deVega:2002im,Derkachov:2002wz,Korchemsky:2003rc} on the proper interpretation for the Odderon solutions  within  this  weak coupling approach.}. The approach of quantum spectral curve~\cite{Alfimov:2014bwa,Gromov:2014bva} holds the promise of further progress in this direction.

We have focused in this study on the leading planar limit. Note that,
in the planar limit, the conformal amplitude growth with a power of
$s$, or, equivalently, $1/\sqrt u$, which would violate the flat-space
Froissart bound. Clearly, in order to address the issue of Froissart
bound for CFT's, one must consider the extension to higher orders in
$1/N^2$. One such re-summation is given by the eikonal approximation.  From the perspective of light-cone OPE, one must begin including multiple-trace primaries in order to carry out such analysis. It is also interesting to examine the effect of confinement deformation. Since scale invariance is broken in the IR, adopting  Poincare-patch for $AdS$  is most suitable for such a treatment.  Instead of spectral curves $\Delta(j)$, one now has ordinary Regge singularities $\alpha(t)$ at positive $t$, leading to discrete physical states at integral $j$, e.g., glueballs. One also finds that our double-Mellin representation, (\ref{eq:newgroupexpansion}) and (\ref{MellinJ}), reduces to a single Mellin (Regge) representation, with a sum over Regge trajectories. These and other related issues will be addressed in future publications.

\section*{Acknowledgements}
We would like to thank B. Basso, M. A. Braun, R. de Mello Koch, V. S. Fadin, V. Gon\c calves, Y. Hatta, G. Levin, G. Korchemsky, Y. V. Kovchegov, L. Mazzucato, B. Nicolescu, M. Paulos and J. Penedones for helpful discussions.  R.B., T.R. and C-I.T.  would like to thank the Physics Department, University of Porto for hospitality during their visits in the summer of 2014 when this work was completed.   The work of R.B. was supported in part by the Department of Energy under Contract No. DEFG02-01ER-40676, that of T.R. and C-I.T. was supported in part by the Department of Energy under Contract No. DE-SC0010010 Task A.
The research leading to these results has received funding from the [European Union] Seventh Framework Programme [FP7-People-2010-IRSES] under grant agreements No 269217, 317089, and from the grant CERN/FP/123599/2011.
\emph{Centro de Fisica do Porto} is partially funded by the Foundation for  Science and Technology of Portugal (FCT). The work of M.D. is supported by the FCT/Marie Curie Welcome II program.

\bibliographystyle{utphys}
\bibliography{conformalodderonDec1}

\end{document}